\documentclass[11pt,a4paper,notitlepage]{article}
\usepackage{amsfonts}
\usepackage[pdftex]{graphics}
\usepackage[a4paper,left=2.5cm,right=2.5cm,top=2cm,bottom=1.5cm,footnotesep=.75cm,headheight=13.6pt]{geometry}
\usepackage{amssymb}
\usepackage{latexsym}
\usepackage[english]{babel}
\usepackage[pdftex,usenames]{color}
\usepackage{eurosym}
\usepackage{hhline}
\usepackage{rotating}
\usepackage[small]{caption}
\usepackage{fancyhdr}
\usepackage[numbers]{natbib}
\usepackage{multirow}
\usepackage{lscape}
\usepackage{geometry}
\usepackage{siunitx}
\usepackage{graphicx}
\usepackage[center]{titlesec}
\usepackage{lipsum}
\usepackage[T1]{fontenc}
\usepackage{array}
\usepackage[utf8]{inputenc}
\usepackage{authblk}
\usepackage{subfigure}
\usepackage{setspace}
\usepackage{graphics}
\usepackage{float}
\usepackage{amsmath}

\providecommand{\abs}[1]{\lvert#1\lvert}

\begin{document}

\title{\vspace{-1cm} Dealing with idiosyncratic cross-correlation when constructing confidence regions for PC factors
\thanks{Financial support from the Spanish National Research Agency (Ministry of Science and Technology) Projects PID2022-139614NB-C21 funded by MCIN/AEI/10.13039/501100011033/FEDER, EU and PID2022-139614NB-C22 is gratefully acknowledged by the first two authors, and the third author, respectively. We are also grateful for useful comments to participants at the XIII$_t$ Workshop in Time Series Econometrics (Zaragoza, March 2023), at the 8th Annual Conference of the Society for Economic Measurement (Milan, July 2023), and at the Computational and Financial Econometrics 2023 meeting (Berlin, December 2023). We are particular indebted to Matteo Barigozzi, Antonio Montañés, and Alexei Onatski for their comments. Obviously, all remaining errors are our only responsibility.}}

\author{Diego Fresoli\\
Dept. of Econ. Analysis: Quantitative Economics, Univ. Autónoma de Madrid, Spain\\
Pilar Poncela\\
Dept. of Econ. Analysis: Quantitative Economics, Univ. Autónoma de Madrid, Spain\\ 
Esther Ruiz\thanks{Corresponding author: Dpto. de Estadística, Universidad Carlos III de Madrid, C/ Madrid 126, 28903 Getafe, Spain, email: ortega@est-econ.uc3m.es, Tel.: +34 916249851}\\
Department of Statistics, Universidad Carlos III Madrid, Spain\\
}
\date{\today}
\maketitle

\maketitle
\begin{abstract}
In this paper, we propose a computationally simple estimator of the asymptotic covariance matrix of the Principal Components (PC) factors valid in the presence of cross-correlated idiosyncratic components. The proposed estimator of the asymptotic Mean Square Error (MSE) of PC factors is based on adaptive thresholding the sample covariances of the idiosyncratic residuals with the threshold based on their individual variances. We compare the finite sample performance of confidence regions for the PC factors obtained using the proposed asymptotic MSE with those of available extant asymptotic and bootstrap regions and show that the former beats all alternative procedures for a wide variety of idiosyncratic cross-correlation structures. 
\end{abstract}



\newpage


\section{Introduction}

Factor extraction is gaining interest among academics, financial markets managers and policy makers when analysing large systems of economic and financial time series. Among the increasing number of applications, one can find the construction of economic or financial indexes, in which the factors have a direct interpretation, and factor-augmented predictive regressions, in which the factors are used to summarize the information contained in a large number of potential predictors; see, for example, Ando and Tsay (2011) and Amburgey and McCracken (2022), in the context of factor-augmented quantile regressions, and Lewis \textit{et al}. (2022) for a very recent application in which they propose building an economic weekly indicator used for prediction of macroeconomic variables. Factors have also been often used to construct scenarios for the underlying economic activity; see, for example, Gonz\'{a}lez-Rivera, Ruiz and Maldonado (2019) and González-Rivera, Rodríguez-Caballero and Ruiz (2024). In these applications, one aims to obtain not only point estimates of the factors but also measures of their uncertainty, which should accompany point estimates in order to establish their statistical legitimacy and allow for the construction of confidence regions; see, for example, Bai (2003, 2004), Bai and Ng (2006), Gon\c calves and Perron (2014, 2020), Aastveit \textit{et al}. (2016), Jackson \textit{et al}. (2016), Gon\c calves, Perron and Djogbenou (2017), Thorsrud (2020), Maldonado and Ruiz (2021), Kim (2022), Barigozzi and Luciani (2023), and Fresoli, Poncela and Ruiz (2023), who argue about the importance of taking into account factor uncertainty in a large range of empirical applications.

Approximate Dynamic Factor Models (DFMs), which are the standard framework for factor extraction, assume that the common variability in a large system of variables is represented by a relatively small number of common factors, while the idiosyncratic components may still have weak cross-correlations; see, for example, Diebold \textit{et al}. (2021), Qiao and Wang (2021), Barigozzi and Luciani (2023), Barigozzi and Lissona (2023) and Gao, Linton and Peng (2024) for empirical applications with cross-correlated idiosyncratic components. 

Within the context of DFMs, factor extraction is often carried out using procedures based on Principal Components (PC), which owe their  popularity to their computational simplicity and well established theoretical properties. PC factors are extracted ignoring the cross-sectional dependence of the idiosyncratic components, with the construction of confidence regions often based on the asymptotic distribution derived by Bai (2003), where the asymptotic Mean Squared Error (MSE) of the factors is often obtained by wrongly assuming cross-sectionally uncorrelated idiosyncratic components.\footnote{It is important to note that the distribution of the factors derived by Bai (2003) refers to the distribution of the factors considered as fixed non-random variables; see Barigozzi and Ruiz (2024).} Although the factors are consistent, wrongly ignoring cross-sectional correlation may have important consequences in the construction of confidence regions.\footnote{Boivin and Ng (2006) show that an excess of amount of cross-sectional correlation among idiosyncratic components worsens the performance of the DFMs; see also Gon\c calves and Perron (2020) and Qiao and Wang (2021).} In a very recent paper, Fresoli, Poncela and Ruiz (2024) carry out Monte Carlo experiments and show that erroneously ignoring cross-correlation of the idiosyncratic components typically lead to wrong confidence intervals for PC factors.

Estimating the MSE of PC factors in the presence of idiosyncratic cross-sectional dependence is a difficult task. Bai and Ng (2006) and Kim (2022) propose taking into account idiosyncratic cross-correlation in obtaining the asymptotic MSE. However, the estimator proposed by Bai and Ng (2006) requires the selection of a subset of cross-sectional variables and there is not a clear criterion about how it should be selected. This selection can be crucial for the performance of the estimator of the MSE and, consequently, for the performance of the confidence regions for the PC factors. On the other hand, the procedure proposed by Kim (2022) relies on the use of a kernel with its performance depending on the choice of the required bandwidth. The bandwidth parameter is chosen by using a wild bootstrap procedure that may involve a heavy computational burden. Alternatively, instead of using the asymptotic distribution, Gon\c calves and Perron (2020) propose constructing confidence regions for the factors using a wild bootstrap procedure that takes into account idiosyncratic cross-correlation. This bootstrap procedure relies on the estimation of the covariance matrix of the idiosyncratic components by only considering the sample covariances that are greater than a given threshold as proposed by Bickel and Levina (2008). The threshold depends on both the cross-sectional and temporal dimensions and is close to zero when both are large. Furthermore, the threshold also depends on a constant, chosen by cross-validation, which is also close to zero when the idiosyncratic cross-sectional dependence, although weak, is large enough.   

The first contribution of this paper is to analyse the performance of extant asymptotic and bootstrap confidence regions for PC factors in the context of approximated DFMs considering several different structures of the idiosyncratic cross-sectional dependence. We show that, when the asymptotic MSE of the factors is estimated ignoring idiosyncratic cross-correlation, the coverages of the corresponding prediction regions can be larger or smaller than the nominal depending on the idiosyncratic correlation structure and the factor loadings. Furthermore, estimating the MSE considering cross-correlation as proposed by Bai and Ng (2006) does not help, unless the structure of the idiosyncratic cross-correlations is known in advance; see also Kim (2020) for the same conclusion. The procedure proposed by Kim (2020) greatly improves the performance of the confidence regions of the factors. Finally, we show that the coverages of bootstrap regions may also be extremely small when the temporal dimension is large and/or the cross-correlation is large enough.

Our second contribution is the proposal of a new estimator of the asymptotic MSE of PC factors that takes into account not only idiosyncratic cross-correlation but also the uncertainty associated with the estimation of the loadings; see Maldonado and Ruiz (2021) for a discussion on the effect of loading estimation on the distribution of the factors. On the one hand, the proposed procedure is similar to that proposed by Kim (2020) but using adaptive thresholding instead of the kernel estimate of the covariance matrix of the idiosyncratic components. Cai and Liu (2011) show that adaptive thresholding is asymptotically valid regardless of the threshold level, which, in this paper, is chosen according to Qiu and Liyanage (2019). By using adaptive thresholding, a specific threshold is used for each entry of the covariances, with the threshold depending on the sample variance of each pairwise sample covariance of the idiosyncratic residuals. The procedure is computationally very simple avoiding the use of bootstrapping or cross-validation to chose the threshold level. On the other hand, once the asymptotic MSE of the factors is obtained, we add an additional term to this MSE to incorporate the uncertainty of the factors due to parameter estimation. This latter correction is based on the subsampling procedure proposed by Maldonado and Ruiz (2021).


The rest of this paper is organized as follows. In Section \ref{sec:DFM}, after describing extant estimators of the asymptotic and bootstrap MSE of PC factors, we propose the new estimator of the asymptotic MSE based on adaptive thresholding. We also describe subsampling procedures designed to take into account parameter uncertainty. In Section \ref{sec:MC}, we report the results of several Monte Carlo experiments designed to analyse the finite sample properties of the confidence intervals of the factors obtained using the proposed procedure and compare them with those of extant procedures. Several structures of the idiosyncratic cross-correlations are considered as well as DFMs with one and two factors. 
Finally, we conclude in Section 5.

\section{Cross-sectionally correlated idiosyncratic components: Point and interval PC extraction of the factors}
\label{sec:DFM}

In this section, after describing extant asymptotic and bootstrap estimators of the MSE of PC factors, we propose a novel asymptotically valid estimator with accurate finite sample coverages in the presence of idiosyncratic cross-correlation. We also describe a subsampling correction of the asymptotic MSE.

\subsection{Confidence regions for principal components factors}

Consider the following stationary approximate DFM
\begin{equation}
\label{eq:DFM}
Y_{t}=\Lambda F_{t}+\text{\ensuremath{\varepsilon_{t}}},
\end{equation}
where $Y_t=\left(y_{1t},...,y_{Nt} \right) ^{\prime}$ is the $N \times 1$ vector of variables observed at time $t$, for $t=1,...,T$, $\Lambda=(\lambda_{1},...,\lambda_{N})^{\prime}$ is the $N\times r$ matrix of loadings, with $\lambda_i$ being the $1 \times r$ vector of loadings of the $r$ factors on the $i$'th variable, $F_{t}$ is the $r\times1$ vector of latent factors, and $\varepsilon_{t}=(\varepsilon_{1t},...,\varepsilon_{Nt})^{\prime}$ is the $N\times1$ vector of idiosyncratic components. The factors are assumed to be stationary. Furthermore, the idiosyncratic components, $\varepsilon_{t}$, are assumed to be white noise with full $N\times N$ variance-covariance matrix given by $\Sigma_{\varepsilon}$, such that the following weak cross-correlation restriction is satisfied
\begin{equation}
l_{max} \left( \Sigma_{\varepsilon} \right) \leq c < \infty, 
\end{equation}
where $l_{max}\left(\Sigma_{\varepsilon} \right)$ is the maximum eigenvalue of $\Sigma_{\varepsilon}$; see, for example, the survey by Stock and Watson (2011). The DFM allowing the idiosyncratic components to be weakly cross-sectionally correlated is known as ``Approximate'' DFM after Chamberlain and Rothschild (1983). Finally, the factors and idiosyncratic components are mutually uncorrelated for all leads and lags and across all units.

On top of the weak cross-correlation restrictions, the DFM in (\ref{eq:DFM}) is assumed to satisfy the assumptions in Bai (2003), needed in the derivation of the asymptotic distribution of the PC factors. Note that, in this paper, we treat the factor loadings as hyperparameters with $\lambda_i$ satisfying $\parallel\lambda_i\parallel = \left[ tr( \lambda_i^{\prime} \lambda_i)\right]^{1/2} \leq M$, where $M$ is a finite positive constant not depending on $N$ and $T$, and the matrix $\Sigma_{\Lambda}=\lim\limits_{N\rightarrow\infty} \frac{1}{N}\Lambda^{\prime}\Lambda$ being positive definite.\footnote{The asymptotic results are the same if the loadings are stochastic, as far as they are independent of the factors and have finite 4th order moments.} Under these assumptions, the factors are pervasive although they are allowed to be temporally dependent.\footnote{Several authors deal with the asymptotic distribution of PC factors in the context of weak factors; see, for example, Bai and Ng (2023) and Uematsu and Yamagata (2023a, 2023b). We conjecture that the presence of idiosyncratic cross-correlations can even be more harmful if the factors are weak.}

Throughout, the number of factors, $r$, is considered as known and fixed with $N$ and $T$. Identification of the loadings and factors in the DFM in (\ref{eq:DFM}) requires imposing $r^2$ restrictions on the parameters. As it is customary in the context of PC factor extraction, we chose to identify factors and loadings by assuming: i) $\frac{F^{\prime} F}{T}=I_r$, where $F$ is the $T \times r$ matrix of factors, and $I_r$ is the $r \times r$ identity matrix; and ii) $\Lambda^{\prime} \Lambda$ is diagonal with distinct elements in the main diagonal arranged in decreasing order.\footnote{Note that to avoid the use of rotations in the asymptotic analysis, these identification restrictions are assumed to be satisfied by the true factors and loadings. Freyaldenhoven (2022) and Fresoli, Poncela and Ruiz (2023) also assume that the restrictions are satisfied when dealing with the asymptotic distribution of the factors. The DFM satisfying the $r^2$ restrictions is called ``identifiable pseudo-true model'' by Jiang, Uematsu and Yamagata (2023), who discuss the asymptotic distribution of the rotated factors when the restrictions are not imposed and there is a rotation matrix that does not depend on data and rotates the true factors to identifiable ones.}

Denote by $Y=\left(Y_1,...,Y_T\right)^{\prime}$, the $T \times N$ matrix of observations. At each moment of time $t$, the PC factors, denoted by $\widetilde{F}_t$, are $\sqrt{T}$ times the eigenvectors corresponding to the $r$ largest eigenvalues of $YY^{\prime}$ arranged in decreasing order. Their corresponding PC loadings are $\widetilde{\Lambda}^{\prime}=\frac{1}{T} \tilde{F}^{\prime}Y$. By construction, $\widetilde{F}$ and $\widetilde{\Lambda}$ are such that $\widetilde{F}^{\prime}\widetilde{F}=I_r$ and $\widetilde{\Lambda}^{\prime}\widetilde{\Lambda}=\widetilde{V}$, where $\widetilde{V}$ is the $r\times r$ diagonal matrix consisting of the first $r$ eigenvalues of the matrix $\frac{1}{TN} Y Y^{\prime}$ arranged in decreasing order. Note that, alternatively, the PC factor, $\tilde{F}_t$, can be expressed as a linear combination of the observations at time $t$ as follows
\begin{equation}
\label{eq:PC}
\tilde{F}_t=\left(\tilde{\Lambda}^{\prime} \tilde{\Lambda} \right)^{-1} \tilde{\Lambda}^{\prime} Y_t.
\end{equation}
Therefore, the factors can be interpreted as the LS estimator with errors in the variables. Stock and Watson (2002) show that, if the cross-sectional correlations of the idiosyncratic noises are weak and the factors are pervasive, the space spanned by the estimated factors is consistent when both $N$ and $T$ tend simultaneously to infinity. Furthermore, Bai (2003) establishes consistency of PC factors in the presence of serial correlation and heteroscedasticity, and derives their asymptotic distribution when both $N, T \rightarrow \infty$ with $\frac{\sqrt{N}}{T}\rightarrow 0$. Imposing the $r^2$ identification restrictions on the factors and loadings described above, Bai and Ng (2013) show that, at each moment of time $t$, this distribution is given by
\begin{equation} \label{eq:asymptotic_1}
\sqrt{N}\left( \widetilde{F}_{t}-F_{t}\right)\stackrel{d} {\rightarrow} N \left( 0,\Sigma _{\Lambda}^{-1}\Gamma _{t}\Sigma _{\Lambda}^{-1}\right),
\end{equation}
where $\Gamma_t = \lim\limits_{N \to \infty} {\frac{1}{N}\sum_{i=1}^N \sum_{j=1}^N \lambda_i^{\prime} \lambda_j E(\varepsilon_{it} \varepsilon_{jt})}$ is a function of the cross-sectional dependence of $\varepsilon_{it}$; see also Barigozzi (2023). Barigozzi and Luciani (2022) show that, if the idiosyncratic noises are serially uncorrelated, the limiting distributions in (\ref{eq:asymptotic_1}) are asymptotically independent across $t$. Finally, it is important to note that asymptotic normality of the factors holds even without assuming normality of the common factors and idiosyncratic components of the DFM. Bai (2003) comments that the restriction $\frac{\sqrt{N}}{T}\rightarrow 0$ is not strong and, therefore, asymptotic normality is the more prevalent situation in empirical applications. 

The asymptotic MSE of $\widetilde{F}_t$ can be estimated as follows
\begin{equation}
\label{eq:avar}
\widehat{Avar}(\widetilde{F}_t)=\frac{1}{N}\left( \frac{\tilde{\Lambda}^{\prime} \tilde{\Lambda}}{N}\right) ^{-1} \widetilde{\Gamma}_t \left( \frac{\tilde{\Lambda}^{\prime} \tilde{\Lambda}}{N}\right) ^{-1},
\end{equation}
where $\widetilde{\Gamma}_t$ is an estimate of $\Gamma_t$. Then, based on the normal asymptotic approximation of the distribution of the factors, one can obtain confidence regions with $100(1-\alpha)\%$ confidence as follows
\begin{equation}
\label{eq:asymregion}
ACR(F_t)=\left\lbrace F_t|\left(F_t-\widetilde{F}_t \right)^{\prime}\left[ \widehat{Avar}(\widetilde{F}_t)\right] ^{-1}\left(F_t-\tilde{F}_t \right) \leq \chi^2_{r}(1-\alpha)\right\rbrace,
\end{equation} 
where $\chi^2_{r}(1-\alpha)$ is the $(1-\alpha)$ quantile of a $\chi^2$ distribution with $r$ degrees of freedom; see Chew (1966).\footnote{As proposed by Lutkephol (1991), confidence regions can also be approximated using Bonferroni rectangles, which are rather popular due to their simplicity.} Note that, if $r=1$ and $\alpha=0.05$, the region in (\ref{eq:asymregion}) reduces to the usual 95\% confidence interval, as follows
\begin{equation}
\label{eq:asyminterval}
ACI(F_t)=[\widetilde{F}_t-1.96 \hat{\sigma}_t^{(F)}, \widetilde{F}_t+1.96 \hat{\sigma}_t^{(F)}],
\end{equation}
where $\hat{\sigma}_t^{(F)}=\sqrt{\widehat{Avar}(\widetilde{F}_t)}$.

\begin{figure}[h!]
\includegraphics[scale=0.25]{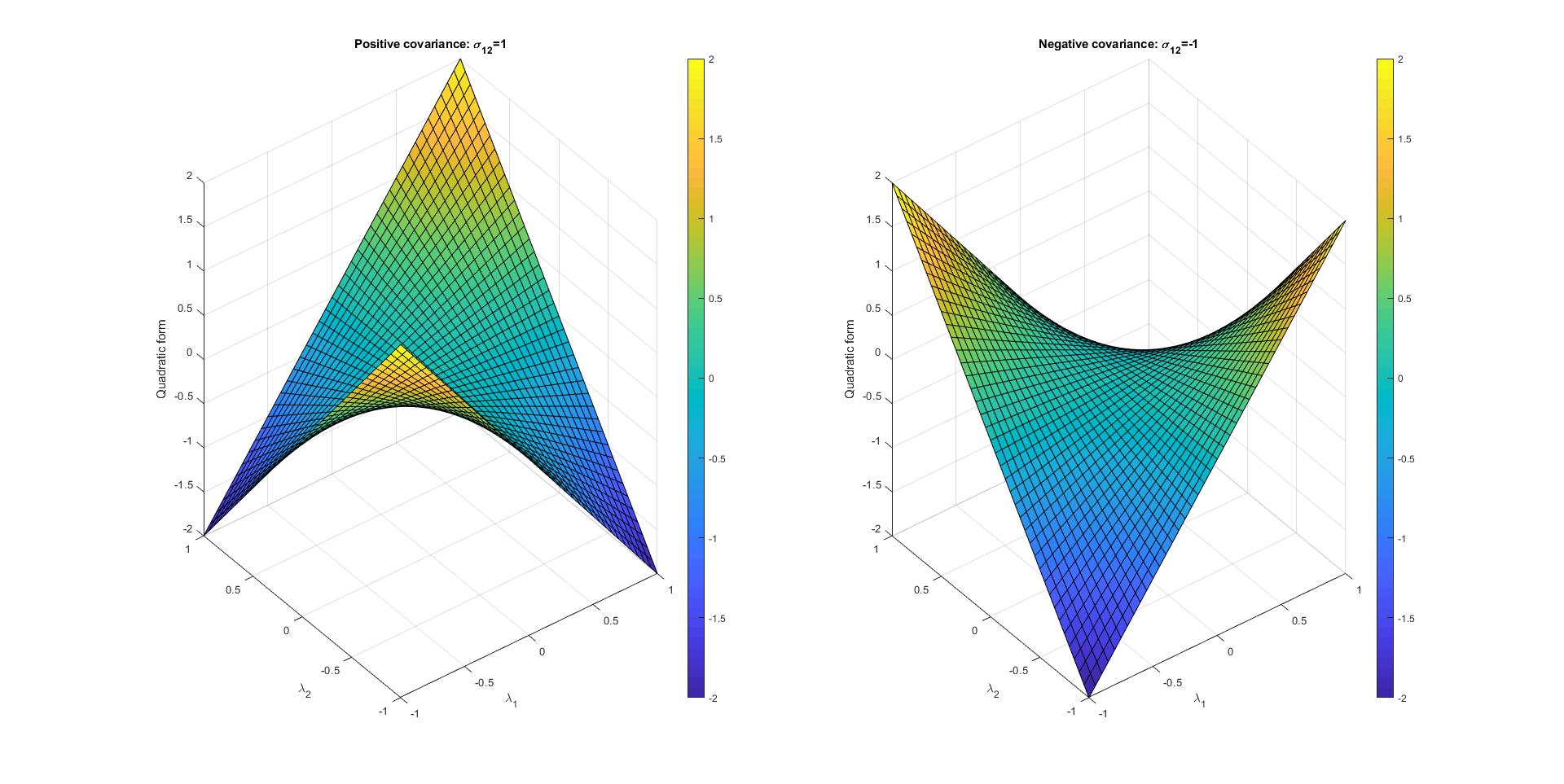}
\caption{Quadratic forms for covariance matrix of PC factors.}
\label{fig:Quadratic}
\end{figure}

The main issue involved in the construction of the confidence regions in (\ref{eq:asymregion}) is the estimation of $\Gamma_t$, which requires that the idiosyncratic cross-sectional dependence is taken into account. In the absence of idiosyncratic cross-sectional correlation, $\Gamma_t$ can be consistently estimated by the following heteroscedasticity robust (HR) estimator
\begin{equation}
\label{eq:gamma_1}
\widetilde{\Gamma}^{HR}_{t}=\frac{1}{N} \widetilde{\Lambda}^{\prime} \widetilde{\Sigma}_{\varepsilon}^* \widetilde{\Lambda}=\frac{1}{N}\sum_{i=1}^{N}\tilde{\lambda}_{i}^{\prime }\tilde{\lambda}_{i}\tilde{\varepsilon}_{it}^{2},
\end{equation}
where $\widetilde{\Sigma}_{\varepsilon}^*$ is a diagonal matrix with $\widetilde{\varepsilon}^2_{it}$ in its main diagonal and $\tilde{\varepsilon}_{it}=y_{it}-\tilde{\lambda}_{i}^{^{\prime}}\widetilde{F_{t}}$ being the PC residuals that estimate the idiosyncratic components. It is important to note that Bai and Ng (2006) propose using the estimator in (\ref{eq:gamma_1}) even in the presence of cross-correlated idiosyncratic components arguing that, if the cross-correlations are small, assuming that they are zero could be convenient because the sampling variability from their estimation could cause an efficiency loss. However, in the presence of non-negligible cross-correlation $\tilde{\Gamma}_t^{HR}$, is inconsistent with the corresponding confidence regions for the factors being invalid. The error of the estimation of the asymptotic MSE of the factors attributed to assuming that $\Sigma_{\varepsilon}$ is diagonal when it is not, depends on $A=\Sigma_{\varepsilon}-\Sigma_{\varepsilon}^*$, where $\Sigma_{\varepsilon}^*=diag(\Sigma_{\varepsilon})$, i.e. the diagonal matrix with the same main diagonal as $\Sigma_{\varepsilon}$. $A$ has zero elements in the main diagonal and non-zero elements in their rows. Consequently, $A$ is a symmetric indefinite matrix and the $r$ quadratic forms involved in $\Lambda^{\prime}A\Lambda$ can take both positive and negative values, depending on the loadings and covariances. Therefore, the asymptotic MSE of the factors computed by wrongly assuming that $\Sigma_{\varepsilon}$ is diagonal, can either overestimate or underestimate the true asymptotic MSE computed with the full $\Sigma_{\varepsilon}$ matrix. As an illustration, Figure \ref{fig:Quadratic} plots quadratic forms $\Lambda^{\prime}A \Lambda$ for the case of a unique factor, $r=1$, and cross-sectional size $N=2$, as functions of the loadings, $\lambda_1$ and $\lambda_2$, when $\sigma_{12}=-1$ and 1. We can observe that $\Lambda^{\prime} \left( \Sigma_{\varepsilon}-\Sigma^*_{\varepsilon} \right) \Lambda$ can be either positive or negative and, consequently, the approximation of the asymptotic MSE of the factor based on $\Sigma^*_{\varepsilon}$ can either underestimate or overestimate the corresponding MSE based on the true $\Sigma_{\varepsilon}$.

The second estimator of $\Gamma_t$ considered takes into account idiosyncratic cross-correlation. It is also due to Bai and Ng (2006), who propose the following cross-sectional HAC (CS-HAC) estimator
\begin{equation} \label{eq:gamma_2}
\widetilde{\Gamma}_{t}^{CS-HAC}=\frac{1}{n}\sum_{i=1}^{n}\sum_{j=1}^{n}\tilde{\lambda}_{i}^{\prime }\tilde{\lambda}_{j} \frac{1}{T} \sum_{t=1}^{T}\tilde{\varepsilon}_{it}\tilde{\varepsilon}_{jt},
\end{equation}
where $n=min[\sqrt{N}, \sqrt{T}]$; see, for example, Ludvigson and Ng (2010) for an application. There are two main issues involved in the CS-HAC estimator. First, although consistent, the CS-HAC estimator requires that the cross-correlation between the idoisyncratic noises is constant over time, i.e. $E(\varepsilon_{it} \varepsilon_{jt})=\sigma_{ij}, \forall t$, which is not necessary for the estimator in (\ref{eq:gamma_1}). In practice, $\widetilde{\Gamma}_t^{CS-HAC}=\widetilde{\Gamma}^{CS-HAC}, \forall t$. Second, an even more important issue is associated with the need of selecting $n$ cross-sectional units, which is required for the convergence of the estimated covariance matrix. The number of units should be such that $\frac{n}{min[N,T]}\rightarrow 0$ when $n\rightarrow \infty$. Without knowing which are the cross-correlated idiosyncratic components, there is no guidance about this selection. Kim (2022) considers two ways of selecting these units. First, one can select $n$ consecutive units. Second, one can select $n$ units randomly.\footnote{Note that the first way of selecting the $n$ units is rather unrealistic as it assumes that the true cross-sectional dependence structure is known with the most correlated units being those which are closest to each other in the cross-section. It could have an interest in applications in which the series in $Y_t$ are grouped together according to their nature. In this case, it could be expected that consecutive series are more correlated than series far apart in the system.} Regardless of whether the first or the second alternative is used, the selection is repeated $G=min[\sqrt{N}, \sqrt{T}]$ times to obtain $\widetilde{\Gamma}^{CS-HAC(g)}, g=1,...,G$, using (\ref{eq:gamma_2}) and, finally, $\widetilde{\Gamma}^{CS-HAC}=\frac{1}{G}\sum_{g=1}^G \widetilde{\Gamma}^{CS-HAC(g)}$. When the consecutive units are selected to compute (\ref{eq:gamma_2}), the estimator is denoted as $\Gamma^{CS-HAC-1}$, while it is denoted as $\Gamma^{CS-HAC-2}$ when the units are randomly selected.

Finally, the third estimator of $\Gamma_t$ considered is due to Kim (2022), who relying on the same stationary assumption described above, proposes the following average sectionally robust estimator
\begin{equation}
\label{eq:kim}
\widetilde{\Gamma}^{AV-SHAC}=\frac{1}{N}\sum_{i=1}^N\sum_{j=1}^Nk\left(\frac{d_{ij}}{d_N} \right) \widetilde{\lambda}_i^{\prime} \widetilde{\lambda}_j \frac{1}{T} \sum_{t=1}^T \widetilde{\varepsilon}_{it} \widetilde{\varepsilon}_{jt},
\end{equation}  
where $k(\cdot)$ is a real-valued kernel function, $d_{ij}$ is a distance between units $i$ and $j$ capturing the strength of the covariance between $\varepsilon_{it}$ and $\varepsilon_{jt}$, and $d_N$ is a bandwidth parameter. Although $\widetilde{\Gamma}^{AV-SHAC}$ depends on the choice of the kernel, $k(\cdot)$, Kim (2022) comments that its finite sample properties are similar regardless of it. In this paper, we use the Parzen kernel. With respect to the distance, Kim (2022) proposes
\begin{equation}
\label{eq:d}
d_{ij}= \left| \frac{1}{Corr(\tilde{\varepsilon}_{it},\tilde{\varepsilon}_{jt})} \right| -1.
\end{equation}
The performance of the estimator in (\ref{eq:kim}) strongly depends on $d_N$. In order to select it, Kim (2022) proposes the following cluster wild bootstrap procedure. First, for $b=1,...,B$, obtain $B=100$ bootstrap replications of the system $Y_t$ as follows
\begin{equation}
\label{eq:resboots}
Y^{*(b)}_t=\tilde{\Lambda} \tilde{F}_t + \varepsilon_t^{*(b)},
\end{equation} 
where the idiosyncratic components are generated by 
\begin{equation}
\label{eq:kimboot}
\varepsilon^{*(b)}_t=\varepsilon^{\dagger(b)}_t \nu_{t}^{(b)},
\end{equation}
where $\varepsilon^{\dagger(b)}_t$ is an $N \times 1$ vector randomly extracted with replacement from $\tilde{\varepsilon}_t$ and $\nu_t^{(b)}$ is an scalar random variable \textit{i.i.d.}(0,1) across $t$. The factors and loadings are estimated by PC from $Y^{*(b)}_t$, obtaining $\tilde{F}_t^{*(b)}$, $\tilde{\Lambda}^{*(b)}$, and the corresponding idiosyncratic residuals $\tilde{\varepsilon}_t^{*(b)}$. For a particular value of the bandwidth, say $d_N^{(1)}$, the corresponding $\widetilde{\Gamma}^{AV-SHAC(b)}(d_N^{(1)})$ is calculated as in (\ref{eq:kim}) with $\tilde{\lambda}_i^{*(b)}$ and $\tilde{\varepsilon}^{*(b)}_{it}$. Finally, compute
\begin{equation}
\label{eq:bootGamma}
T(d_N^{(1)})=\frac{1}{B} \sum_{b=1}^B \left(\tilde{\Lambda}^{\prime} \tilde{\Lambda} \right) ^{-1} \widetilde{\Gamma}^{AV-SHAC(b)}(d_N^{(1)}) \left(\tilde{\Lambda}^{\prime} \tilde{\Lambda} \right) ^{-1}.
\end{equation}

The procedure above is repeated for $M$ values of the bandwidth parameter within $D=(d_N^{(1)},...,d_N^{(M)})$, which is a set of reasonable values for $d_N$ for a given sample size. In practice, we follow Kim (2022) and implement the wild bootstrap procedure above for $d_N=(0.5, 1.0, 1.5,...,20)$. Second, for each $d_N^{(i)}$, $i=1,...,M$, compute the average number of pseudo-neighbours given by $l(d_N^{(1)})=\frac{1}{N} \sum_{i=1}^N \sum_{j=1}^N I(d_{ij}\leq d_N^{(1)})$, where $I(\cdot)$ is an indicator function that takes value 1 if the argument is true and zero otherwise. The value of $d_N$ is chosen by the following optimization
\begin{equation}
\label{eq:opt}
d_N^{\dagger} = {arg\max_{d_N\in D_N}} T(d_N)
\end{equation}
\begin{equation*}
\textrm{s.t. } l(d_N) \leq c_{\pi} min[N,T]^{\pi},
\end{equation*}
where, as suggested by Kim (2022), we chose $\pi=2/3$ and $c_{\pi}=1$.

Note that the wild bootstrap procedure to select the bandwidth is very demanding from a computational point of view. 

\subsection{Taking into account estimation of the loadings: Subsampling}

The convergence rate of PC factors, $min\{\sqrt{N}, T\}$, reflects the fact that the factor loadings are unknown and have to be estimated. As pointed out in (\ref{eq:PC}), if the factor loadings in $\Lambda$ were known, $F_t$ can be estimated by using LS in each cross-section. In this case, the rate of convergence is $\sqrt{N}$. However, in practice, the loadings are unknown. Given that the individual loadings can be estimated no faster than $O_p(T^{-1/2})$, the asymptotic distribution in (\ref{eq:asymptotic_1}) requires a relatively large $T$. But large sample sizes are not always available; see the discussion by Fan, Liao and Wang (2016). When $T$ is not large enough, the asymptotic distribution could not be a good approximation to the finite sample distribution of the estimated factors.

To solve this issue, Maldonado and Ruiz (2021) propose a correction of the finite sample approximation of the asymptotic covariance matrix of the factors designed to measure the uncertainty associated with the estimation of the factor loadings. The proposed correction is based on subsampling subsets of series in the cross-sectional space, with each series containing all temporal observations.

Although the subsampling correction was proposed by Maldonado and Ruiz (2021) in the context of the HR estimator of the MSE of the factors, it is important to note that it can be used regardless of the particular estimator of $\Gamma$ implemented. The corresponding estimators will be denoted as HR*, SC-HAC1*, SC-HAC2*, and AV-SHAC*. 

\subsection{Bootstrap distribution of the factors}

Instead of using the asymptotic distribution to construct confidence regions for the factors, one can use a bootstrap distribution. Several bootstrap procedures have been proposed in the context of DFMs although most of them are designed for other objectives than constructing confidence intervals for the factors; see, for example, the discussion by Maldonado and Ruiz (2021). Early on, Ludvigson and Ng (2007, 2009, 2010) propose obtaining $B$ bootstrap replicates of $Y_t$ as in (\ref{eq:resboots}) with $\varepsilon_t^{*(b)}$ being random extractions with replacement from $\tilde{G}_{\varepsilon}$, the empirical distribution of the vector of centred PC residuals $\tilde{\varepsilon}_t-\bar{\tilde{\varepsilon}}=(\tilde{\varepsilon}_{1t}-\bar{\tilde{\varepsilon}}_1,...,\tilde{\varepsilon}_{Nt}-\bar{\tilde{\varepsilon}}_N)^{\prime}$, where $\bar{\tilde{\varepsilon}}_{i}=\frac{1}{T}\sum_{t=1}^T \tilde{\varepsilon}_{it}$.\footnote{The replications obtained in (\ref{eq:resboots}) do not reproduce the uncertainty associated with the estimation of the loadings.} Note that each vector resampled from $\tilde{\varepsilon}_t-\bar{\tilde{\varepsilon}}$ contains all the idiosyncratic components and, consequently, preserves their cross-sectional dependence. Gon\c calves (2011) shows the validity of (\ref{eq:resboots}) for inference in a linear panel data model with fixed effects. Alternatively, instead of randomly extracting vectors from $\tilde{G}_{\varepsilon}$, Gon\c calves, Perron and Djogbenou (2017) propose using the following wild bootstrap procedure
\begin{equation}
\label{eq:boot1}
\varepsilon_{it}^{*(b)}=\left( \tilde{\varepsilon}_{it} - \bar{\tilde{\varepsilon}}_i \right) \nu_{it}^{(b)},
\end{equation}
with $\nu^{(b)}_{it}$ being $i.i.d.(0,1)$ across $i$ and $t$. Assuming lack of idiosyncratic cross-sectional correlation, Gon\c calves, Perron and Djogbenou (2017) justify using the wild bootstrap to consistently estimate the distribution of PC factors. It is important to note that, regardless of whether the $i.i.d.$ or the wild bootstrap procedure is implemented, Gon\c calves and Perron (2020) show that resampling cross-sectional vectors over time is invalid to estimate the asymptotic MSE of the factors. In fact, using either of both bootstrap procedures, it is easy to see that $E^*(\varepsilon_t^{*})=0$ and $Var^*(\varepsilon_t^*)=\frac{1}{T}E^*(\varepsilon_t^* \varepsilon_t^{*\prime})=\frac{1}{T}\sum_{t=1}^T (\tilde{\varepsilon}_t-\bar{\tilde{\varepsilon}})(\tilde{\varepsilon}_t-\bar{\tilde{\varepsilon}})^{\prime}=\frac{\widetilde{\varepsilon}^{\prime} \widetilde{\varepsilon}}{T}-\frac{\widetilde{\varepsilon}^{\prime} \iota \iota^{\prime} \widetilde{\varepsilon}}{T^2}$, where $\iota$ is a $T \times 1$ vector of ones. The first order conditions for the estimation of the PC factors imply that $\widetilde{\Lambda}^{\prime}\widetilde{\varepsilon}^{\prime}_t=0$. Consequently,
\begin{equation}
\label{eq:gamma_boot}
\Gamma^*=\frac{1}{T}\sum_{t=1}^T \frac{1}{N} \tilde{\Lambda}^{\prime} Var^*(\varepsilon_t^*) \tilde{\Lambda}=0.
\end{equation}

More recently, Gon\c calves and Perron (2020) propose the cross-sectional dependent (CSD) bootstrap, which takes into account the presence of idiosyncratic cross-correlation. Since $\Gamma$ is a function of the cross-sectional dependence of $\varepsilon_{it}$, $\varepsilon^*_{it}$ should be chosen in a way that it replicates this cross-correlation (and cross-sectional heteroscedasticity, if it exists). In fact, one should choose $\varepsilon_t^*$ such that
\begin{equation}
\Gamma^*=\frac{1}{T}\sum_{t=1}^T Var^{*}\left(\frac{1}{\sqrt{N}} \widetilde{\Lambda}^{\prime} \varepsilon_t^* \right) \xrightarrow{p} \Gamma. 
\end{equation}
Gon\c calves and Perron (2020) propose obtaining bootstrap samples as follows
\begin{equation}
\varepsilon^{*(b)}_t=\widehat{\Sigma}_{\varepsilon}^{1/2} \nu^{(b)}_t,
\end{equation}
where $\nu^{(b)}_t=(\nu_{1t}^{(b)}, \nu_{2t}^{(b)},...,\nu_{Nt}^{(b)})$ is defined as in (\ref{eq:boot1}) and $\widehat{\Sigma}_{\varepsilon}$ is estimated by the thresholding estimator proposed by Bickel and Levina (2008), which is based on keeping only those covariances between $\tilde{\varepsilon}_{i}$ and $\tilde{\varepsilon}_j$ over a given threshold (universal hard thresholding rule), as follows\footnote{$\widetilde{\Sigma}_{\varepsilon}=\frac{1}{T} \sum_{t=1}^T \tilde{\varepsilon}_t \tilde{\varepsilon}_t^{\prime}$ is not a good choice as it is not consistent in the spectral norm and implies $\Gamma^*=0$, as in the case of the $i.i.d.$ bootstrap.}
\begin{equation}
\label{eq:threshold}
\end{equation}
where $\tilde{\sigma}_{ij}=\frac{1}{T} \sum_{t=1}^T \left(\tilde{\varepsilon}_{it}- \bar{\tilde{\varepsilon}}_i \right) \left(\tilde{\varepsilon}_{jt}- \bar{\tilde{\varepsilon}}_j \right)$ and $\omega=c \left( \frac{1}{\sqrt{N}} + \sqrt{\frac{log (N)}{T}} \right)$, with $c>0$ being a sufficiently large constant chosen by cross-validation with an upper bound restriction imposed on the maximum number of non-zero elements of $\widehat{\Sigma}_{\varepsilon}$ across rows, namely, $m_N\equiv{\max}_{i\leq N} \sum_{j=1}^N 1(\hat{\sigma}_{ij}\neq 0)$ such that $m_N=o(min(\sqrt{N}, \sqrt{\frac{T}{log(N)}}))$.\footnote{In their Monte Carlo simulations, Gon\c calves and Perron (2020) fix $m_N=11$.}

After obtaining $B$ replications of $Y_t$ as in (\ref{eq:resboots}), the factors are extracted from each replication using PC, obtaining $\tilde{F}_t^{*(1)}, \tilde{F}_t^{*(2)},...,\tilde{F}_t^{*(B)}$. Their empirical bootstrap distribution can be used to obtain the corresponding bootstrap ellipsoid with confidence $100(1-\alpha)\%$ as follows
\begin{equation}
\label{eq:bootregion}
BCE(F_t)=\left\lbrace F_t | \left[F_t - \bar{F}^*_t \right]^{\prime} S_{F^*}^{-1} \left[F_t - \bar{F}^*_t \right] < Q_{\delta}^* \right\rbrace 
\end{equation}
where $\bar{F}^*_t$ is the sample mean of the $B$ bootstrap replicates, $S_{F^*}$ is the corresponding sample covariance and $Q_{\delta}^*$ is the $100(1-\alpha)\%$ percentile of the empirical bootstrap distribution of the quadratic form $\left[F_t - \bar{F}^*_t \right]^{\prime} S_{F^*}^{-1} \left[F_t - \bar{F}^*_t \right]$. Obviously, one can also obtain individual intervals for each of the factors.

\begin{sidewaysfigure}
\begin{centering}
\includegraphics[scale=0.23]{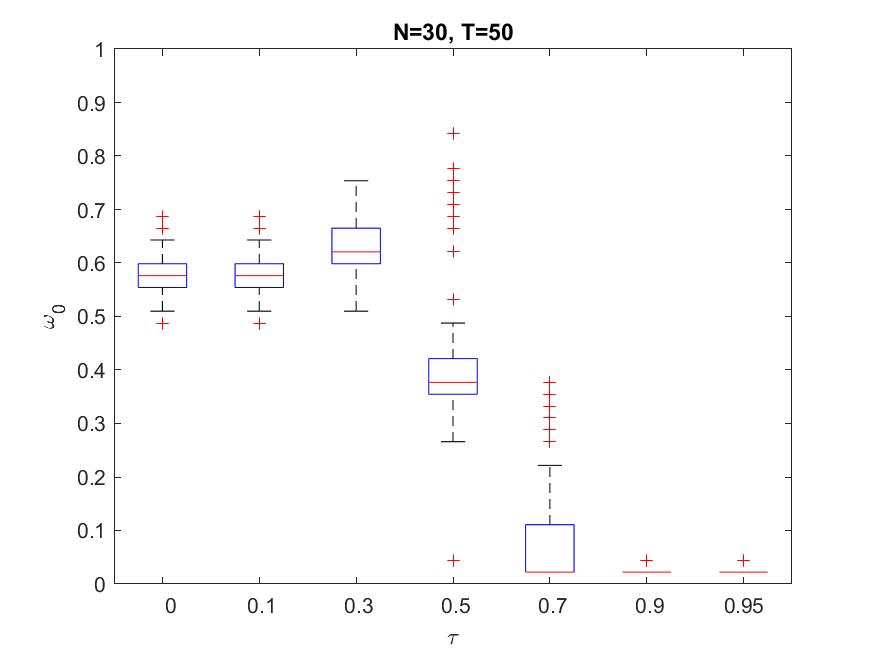}
\includegraphics[scale=0.23]{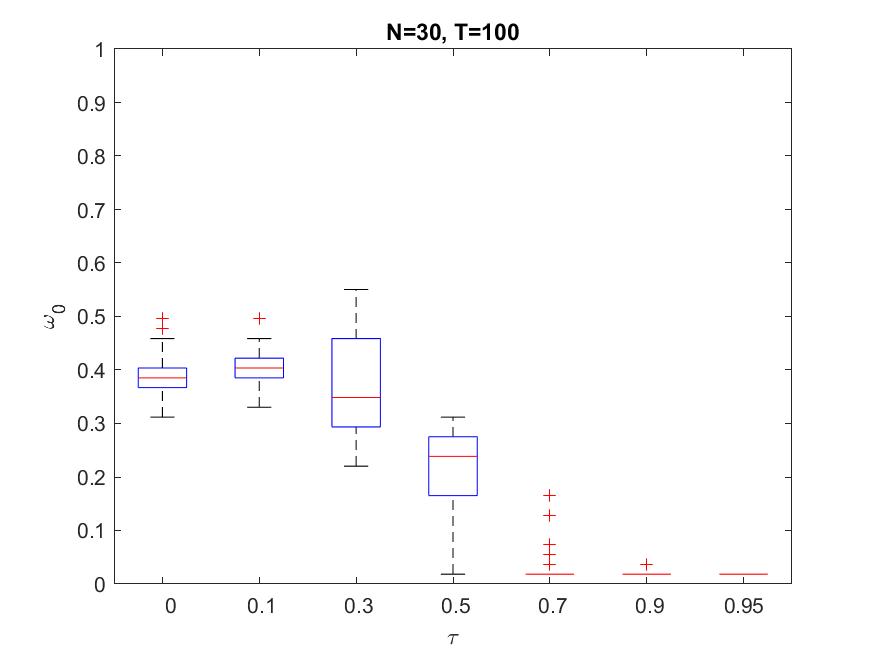}
\includegraphics[scale=0.23]{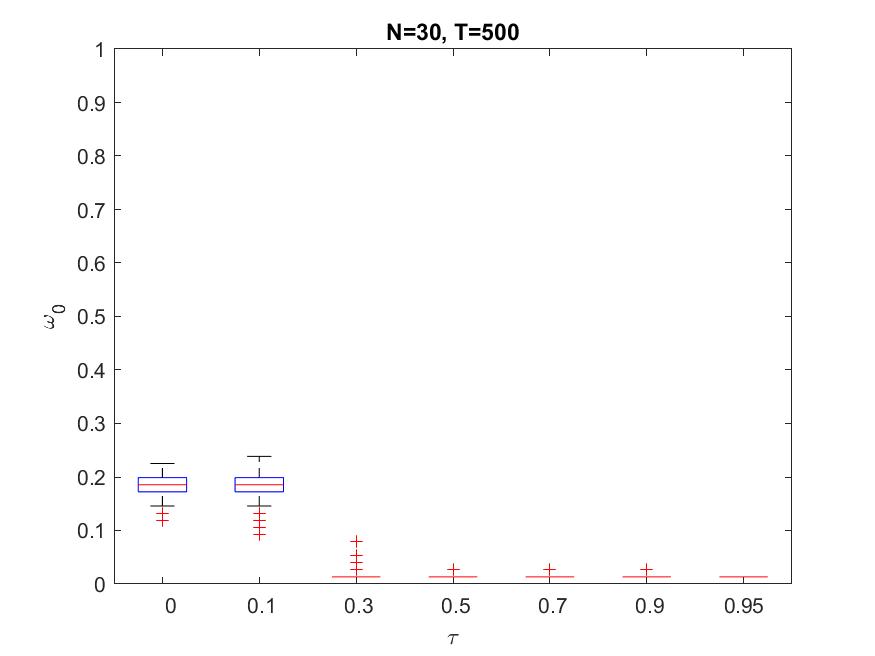}
\par\end{centering}
\begin{centering}
\includegraphics[scale=0.23]{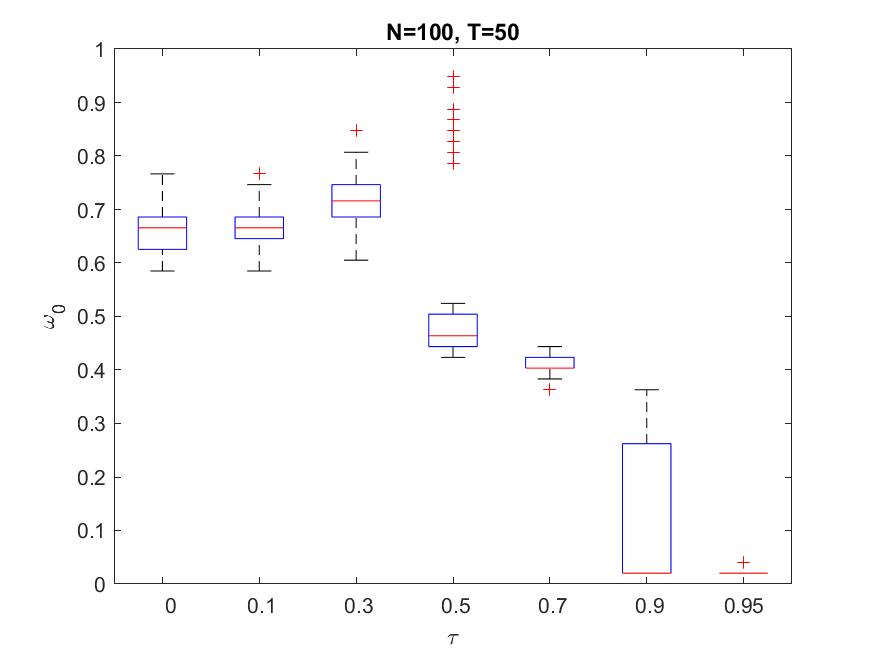}
\includegraphics[scale=0.23]{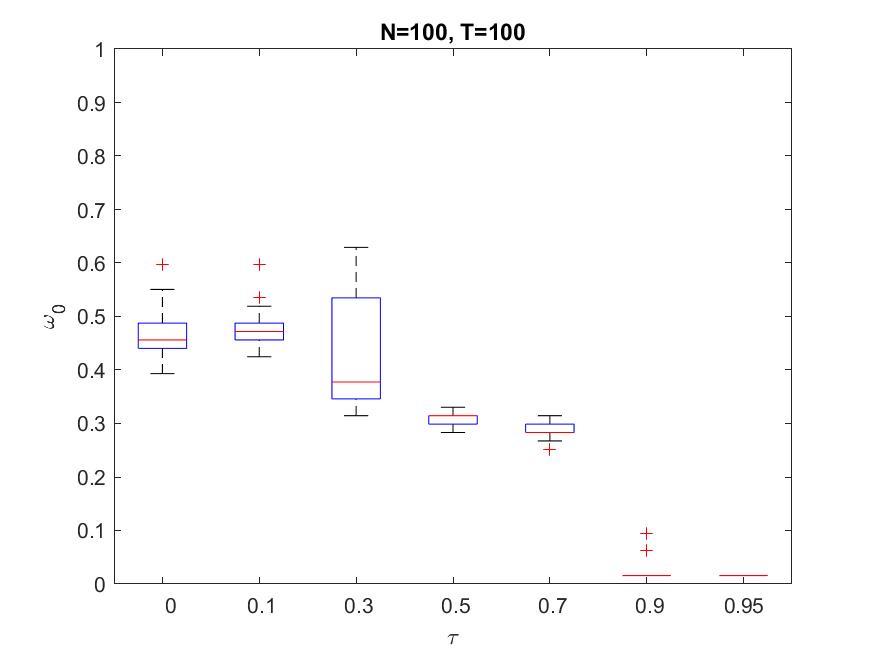}
\includegraphics[scale=0.23]{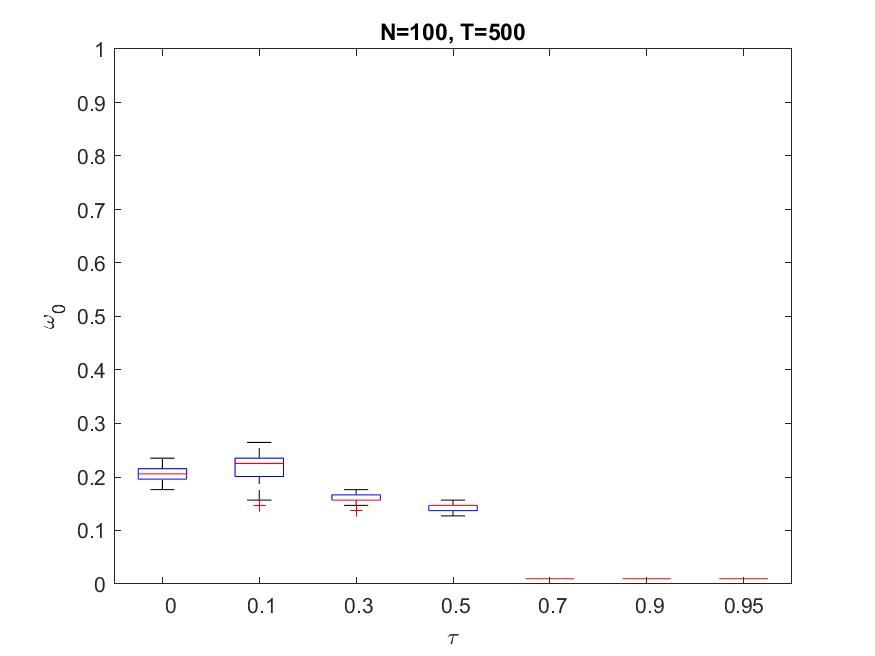}
\par\end{centering}
\begin{centering}
\includegraphics[scale=0.23]{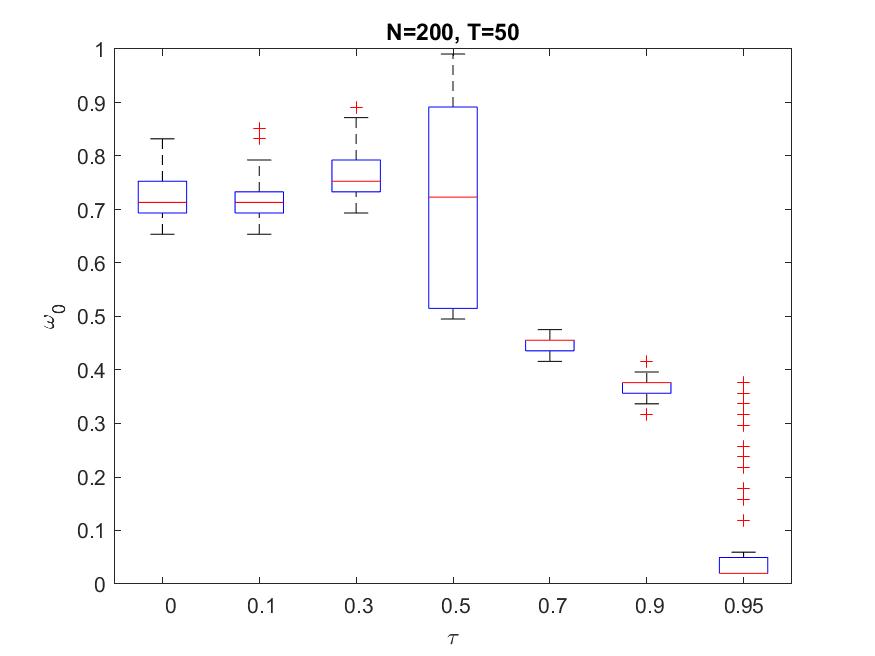}
\includegraphics[scale=0.23]{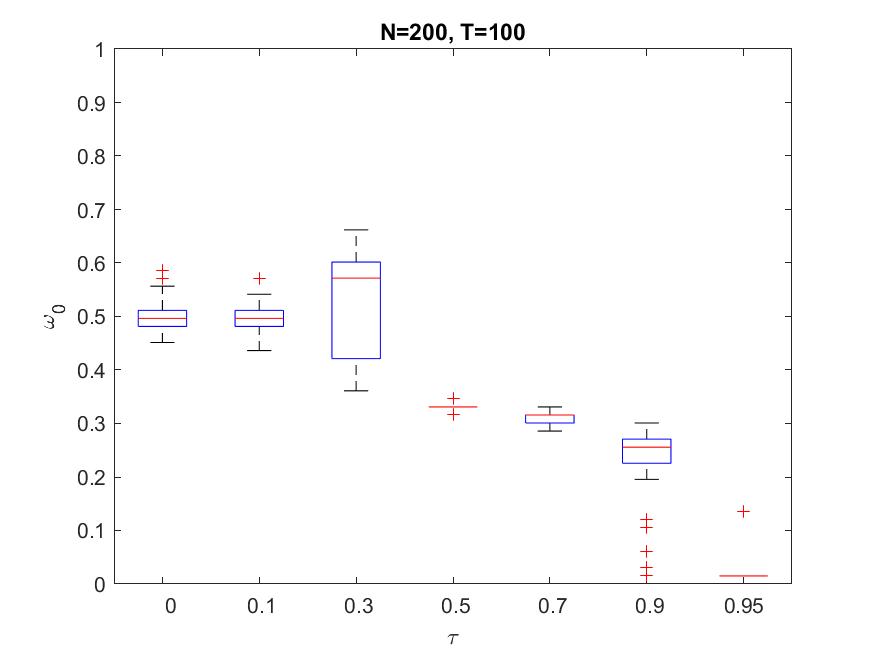}
\includegraphics[scale=0.23]{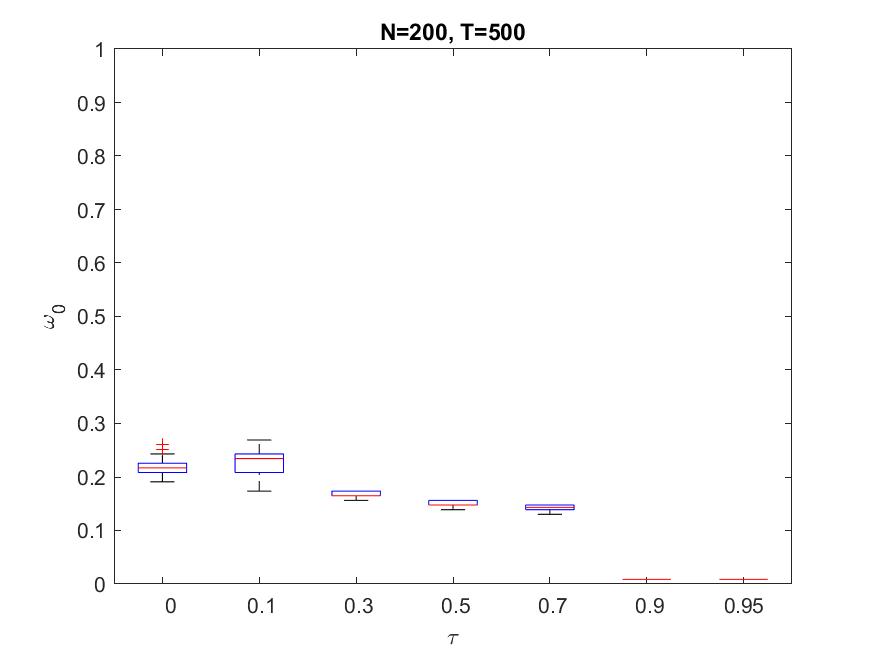}
\par\end{centering}
\caption{Threshold $\omega$ for sample autocovariances when the cross-correlation structure is Toeplitz with positive $\tau$ parameter.}
\label{fig:Thresh}
\end{sidewaysfigure}

It is important to note that the thresholding assumed in (\ref{eq:threshold}) implies that, when $T$ and $N$ go to $\infty$, with $T$ going faster than $N$, the threshold $\omega=0$. Furthermore, the constant $c$ chosen by cross-validation is zero when the cross-sectional dependence of the idiosyncratic noises is large enough (even if the assumption of weak cross-correlation is satisfied); see Wang and Liu (2017), who put forward the drawbacks of cross-validation in high dimensional data analysis. To illustrate this point, Figure \ref{fig:Thresh} plots box-plots of the selected threshold $\omega$ in a simulation with 1000 systems of variables with cross-sectional dimension $N=30, 100$ and 200 variables and $T=50, 100$ and 500 observations, generated by model (\ref{eq:DFM}) with $r=1$ factor, and with the idiosyncratic components, $\varepsilon_t$, generated by a multivariate Gaussian white noise with covariance matrix $\Sigma_{\varepsilon}$ with the elements in the main diagonal of $\Sigma_{\varepsilon}$ generated by $\sigma_i^2\sim U(0.5, 10)$ while the off-diagonal elements are generated according to the following Toeplitz structure 
\begin{equation}
\label{eq:cross}
\sigma_{ij}=\sigma_i \sigma_j \tau^{\abs{i-j}}, i=1,...,N, j=i+1,...,N,
\end{equation}
with $\tau=0, 0.1, 0.3, 0.5, 0.7, 0.9$ and 0.95. Figure \ref{fig:Thresh} shows that, regardless of the true magnitude of the idiosyncratic cross-correlations, $\tau$, the threshold for the sample covariances is close to zero when $T$ is large.\footnote{Note that the sample correlations are approximately given by the covariances divided by 5.} If $\tau$ is large, the threshold is such that only very small sample cross-correlations are trimmed. Figure \ref{fig:Thresh} also shows that, in the context of the particular structure of the covariances considered, only for small values of $\tau$ and large cross-sectional dimension, $N$, there is some relevant trimming of the sample covariances. Finally, note that, when the threshold is zero $\widehat{\Sigma}_{\varepsilon}=\widetilde{\Sigma}_{\varepsilon}$, the sample covariance matrix of the idiosyncratic residuals. In this later case, Gon\c calves and Perron (2020) show that $\Gamma^*$ is given by (\ref{eq:gamma_boot}) and, therefore, is equal to zero. 

\subsection{A new procedure for estimating the asymptotic MSE of the factors in the presence of idiosyncratic cross-correlation}

Extant procedures to construct confidence regions for PC factors may have difficulties in the presence of idiosyncratic cross-sectional dependence. As described above, when confidence regions are based on the asymptotic distribution with the MSE computed by wrongly assuming that the idiosyncratic components are cross-sectionally uncorrelated, their coverages can be either larger or smaller than the nominal; see Fresoli, Poncela and Ruiz (2023), who show that, when the structure of the cross-correlations is represented by a Toeptliz matrix with positive parameter, the coverages are typically smaller than the nominal. On the other hand, the available estimators of the asymptotic MSE require either to select a number of cross-sectional units or a bandwidth parameter. In any case, the performance of the estimator heavily relies on this choice. Finally, when, as proposed by  Gon\c calves and Perron (2020), the confidence regions are obtained using the bootstrap distribution, the use of the thresholding by Bickel and Levina (2008) may be problematic for large levels of cross-correlation.

Consequently, in this paper, we propose a novel computationally simple estimator of the asymptotic covariance matrix of PC factors, based on adaptive thresholding of the sample covariances of the idiosyncratic residuals. The adaptive thresholding procedure, proposed by Cai and Liu (2011), is based on the variance of each individual entry of the sample covariances, resulting in individual specific threshold levels for each entry, which leads to a more accurate covariance estimator when compared to the thresholding proposed by Bickel and Levina (2008). Furthermore, the estimator proposed in this section does not need choosing a particular threshold avoiding the problems associated to using cross-validation. Finally, the proposed estimator is consistent.

Assuming that the pairwise cross-correlations between the idiosyncratic components are constant over time, the proposed cross-sectionally robust White-type estimator of $\Gamma$ is as follows
\begin{equation}
\label{eq:Gamma_new}
\widetilde{\Gamma}^{AT-CSR}= \frac{1}{N} \sum_{i=1}^N \sum_{j=1}^N \tilde{\lambda}^{\prime}_i \tilde{\lambda}_j \frac{1}{T} \sum_{t=1}^T \tilde{\varepsilon}_{it} \tilde{\varepsilon}_{jt} I\left(\mid \tilde{\sigma}_{ij}\mid \geq c_{ij} \right), 
\end{equation}
where $\tilde{\sigma}_{ij}$ is the sample covariance of the idiosyncratic residuals defined as in (\ref{eq:threshold}). The threshold $c_{ij}$ is defined as proposed by Cai and Liu (2011) by taking into account the variance of $\tilde{\sigma}_{ij}$ as follows
\begin{equation}
c_{ij}=\delta \left[ \widehat{Var}\left[\tilde{\varepsilon}_{it} \tilde{\varepsilon}_{jt} \right] \frac{log(N)}{T}  \right] ^{1/2},
\end{equation}
where $\widehat{Var}\left[\tilde{\varepsilon}_{it} \tilde{\varepsilon}_{jt} \right]= \frac{1}{T}\sum_{t=1}^T \left[\left(\tilde{\varepsilon}_{it} - \bar{\tilde{\varepsilon}}_i \right) \left(\tilde{\varepsilon}_{jt} - \bar{\tilde{\varepsilon}}_j \right) - \tilde{\sigma}_{ij} \right] ^2$, and $\delta$ is the threshold level. Qiu and Liyanage (2019) propose the following optimal threshold level that minimizes the Frobenius risk
\begin{equation}
\delta=\left[ 2(2-\gamma)\right] ^{1/2},
\end{equation}
where $\gamma \in (0,2)$ depends on the sparsity of $\Sigma_{\varepsilon}$ with $\gamma=0$ representing a highly sparse matrix while $\gamma=2$ represents a non-sparse covariance matrix of the idiosyncratic components. Cai and Liu (2011) show that the value of $\delta$ can be taken as fixed or chosen empirically by cross-validation with the estimator of the covariance matrix attaining the optimal rate of convergence under the spectral norm in both cases. Although the choice of $\delta$ will not affect the rate of convergence, it could affect the numerical performance of the resulting estimator.

It is important to mention that the estimators of the covariance matrix based on thresholding do not guarantee its positive definiteness. Consequently, as in Goncalves and Perron (2020) and Kim (2022), we implement the correction suggested by Politis (2011) to render it positive definite. In particular, if the estimated covariance matrix, say $S$, is not positive definite, we obtain a corrected matrix $B=AD+A$, with $A$ being the matrix of the orthonormal eigenvectors of $S$, and $D=dig(a_1^+,...,a_N^+)$, where $a_i^+=max(a_i,c)$ and $a_i$ are the eigenvalues of $S$.  

After estimating the asymptotic MSE using (\ref{eq:gamma_1}) with $\hat{\Gamma}_t$ estimated as in (\ref{eq:Gamma_new}), we propose to correct it by using subsampling to incorporate the uncertainty due to loading estimation. The resulting estimator will be denoted as AT-CSR*.

\section{Monte Carlo experiments}
\label{sec:MC}

In this section, we analyse the finite sample coverages of the confidence intervals/regions based on the proposed estimator of the asymptotic MSE of PC factors and compare them with those of extant asymptotic and bootstrap regions/intervals. We also analyse the performance of the subsampling procedure designed to deal with parameter uncertainty. 
We consider designs with structures of the correlation matrix of the idiosyncratic noises with negative correlations and/or without a Toeplitz structure, and with two factors.

\subsection{Monte Carlo design}

\begin{sidewaysfigure}
 {\hspace{1.4cm} $\tau=0.00$} {\hspace{2.0cm} $\tau=0.10$} {\hspace{2.0cm} $\tau=0.30$}  {\hspace{2.0cm} $\tau=0.50$} {\hspace{2.0cm} $\tau=0.70$}  {\hspace{2.0cm} $\tau=0.90$} {\hspace{2.0cm} $\tau=0.95$}

\smallskip{}

\rotatebox{90}{\hspace{0.5cm} Toeplitz pos.  $\tau$  \hspace{0.5cm}}$\,\,$\includegraphics[width=3.5cm,height=3.5cm]{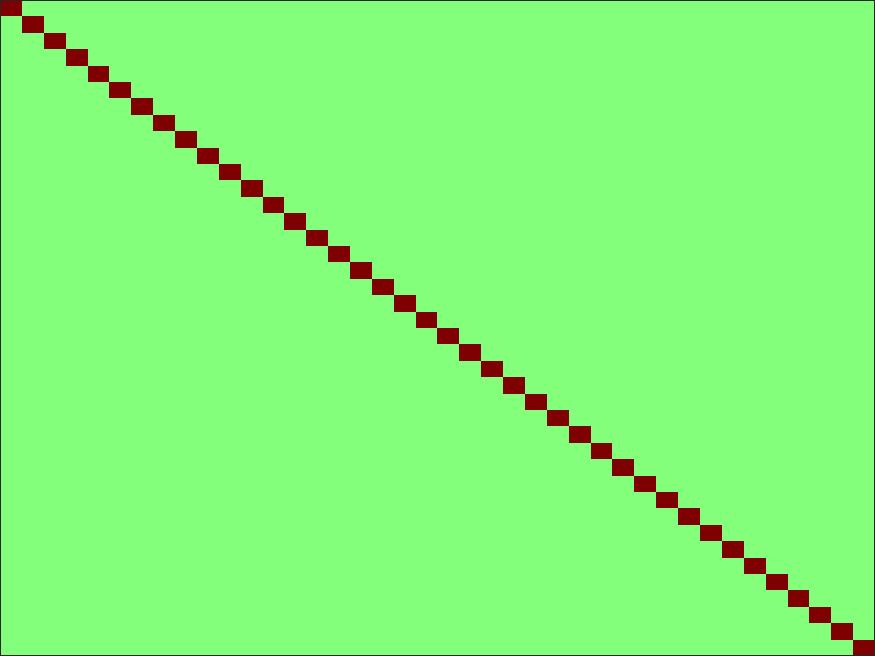}\includegraphics[width=3.5cm,height=3.5cm]{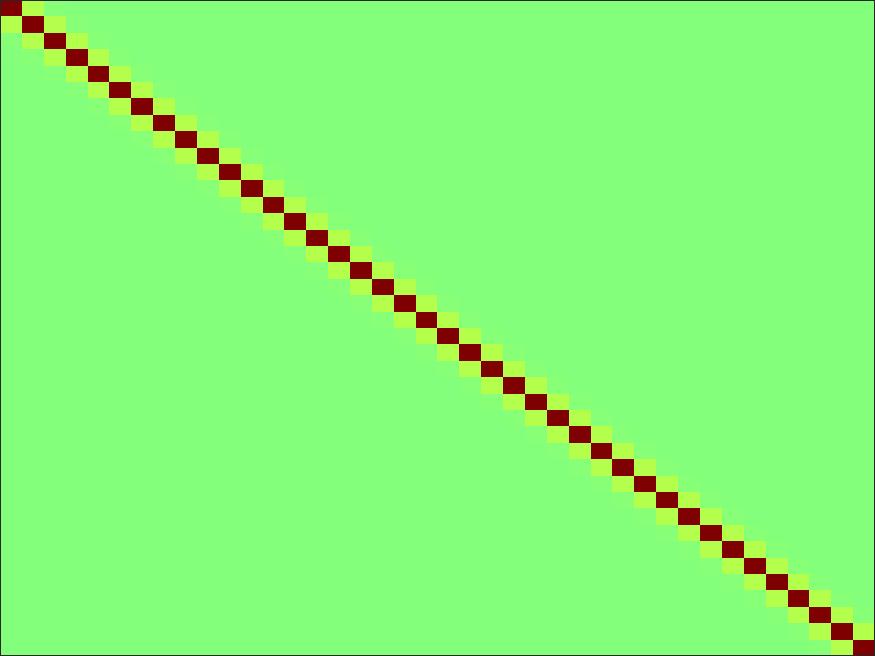}\includegraphics[width=3.5cm,height=3.5cm]{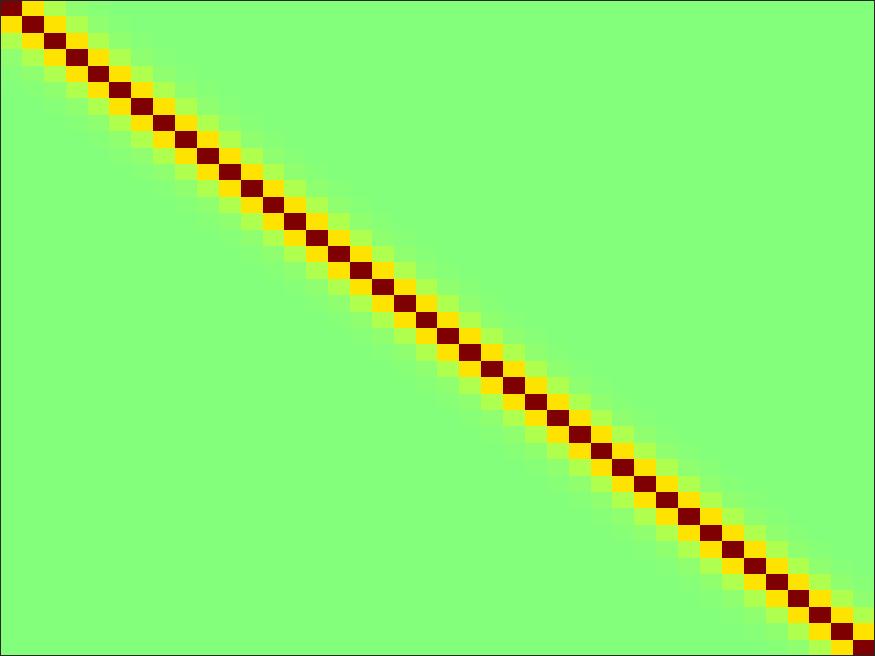}\includegraphics[width=3.5cm,height=3.5cm]{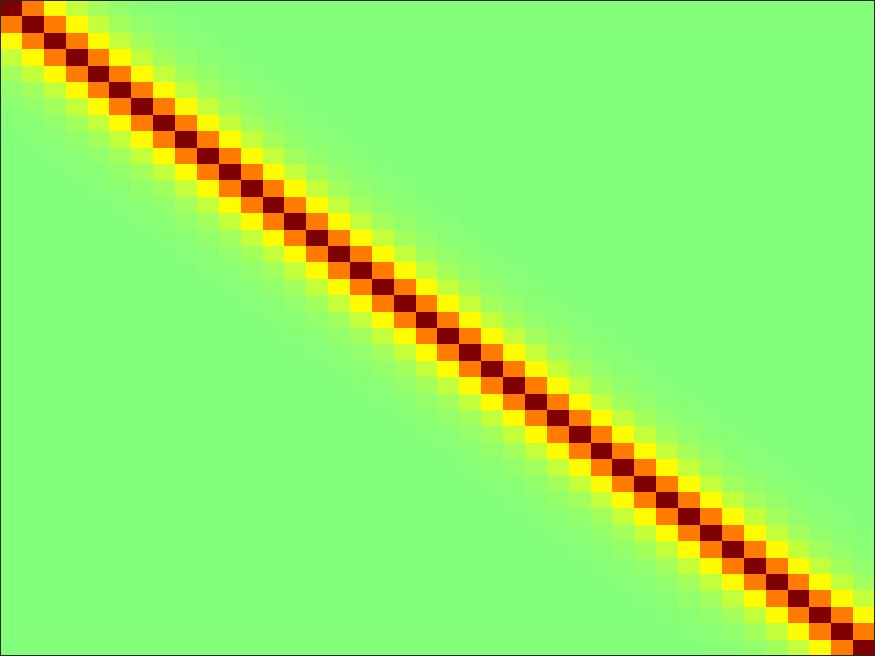}
\includegraphics[width=3.5cm,height=3.5cm]{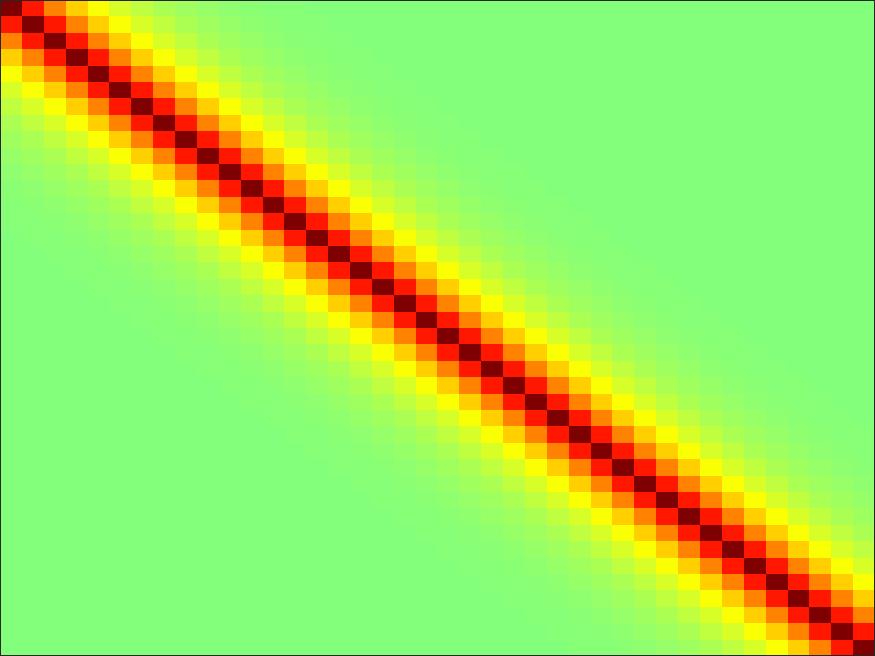}\includegraphics[width=3.5cm,height=3.5cm]{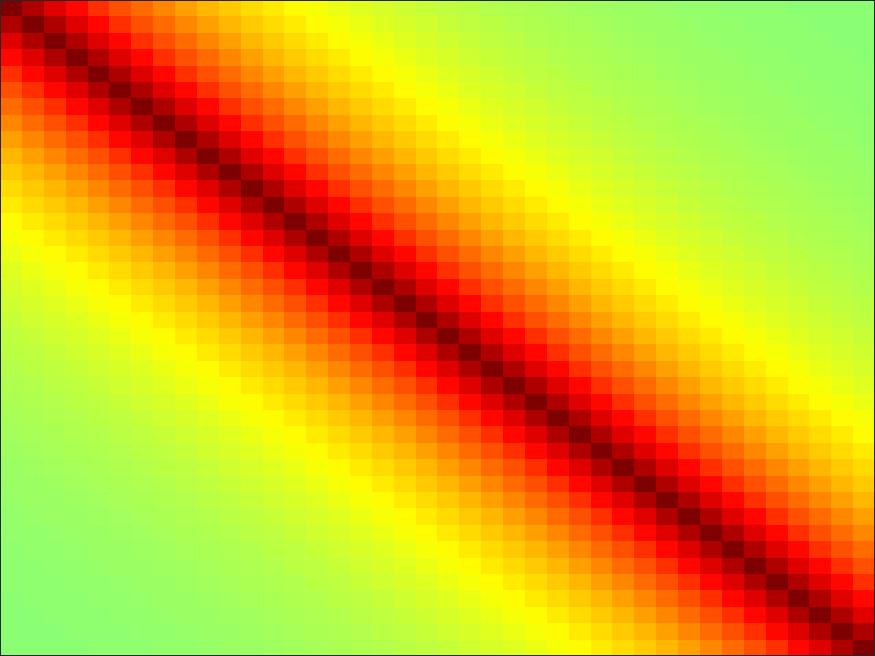}\includegraphics[width=3.5cm,height=3.5cm]{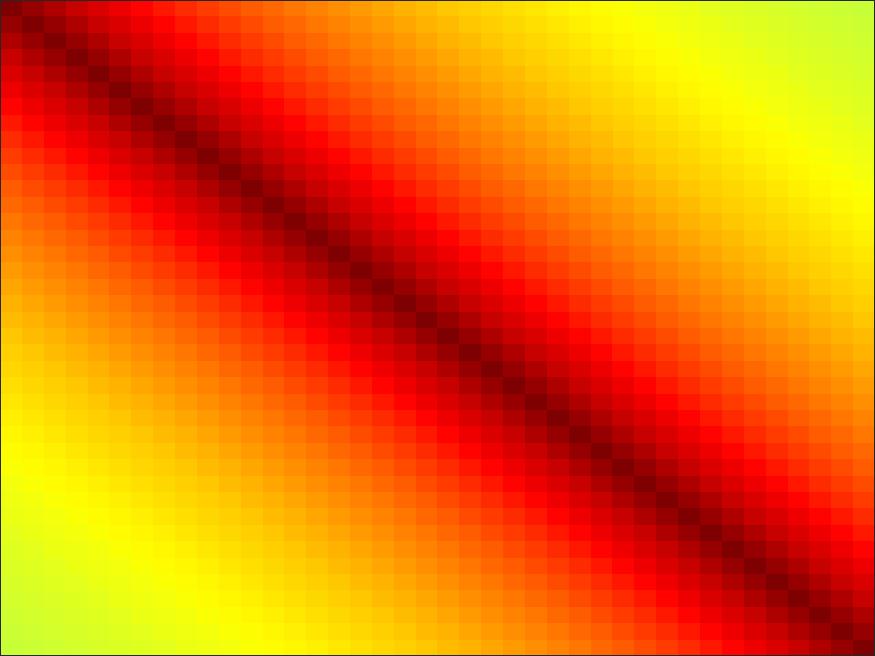}\medskip{}

\rotatebox{90}{\hspace{0.5cm}  Permuted   \hspace{0.5cm}}$\,\,$\includegraphics[width=3.5cm,height=3.5cm]{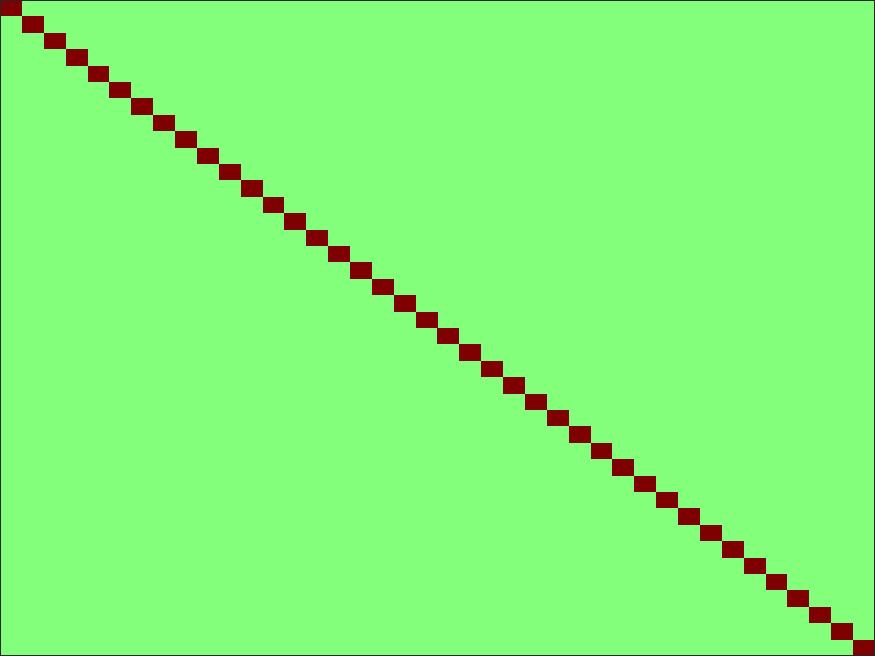}\includegraphics[width=3.5cm,height=3.5cm]{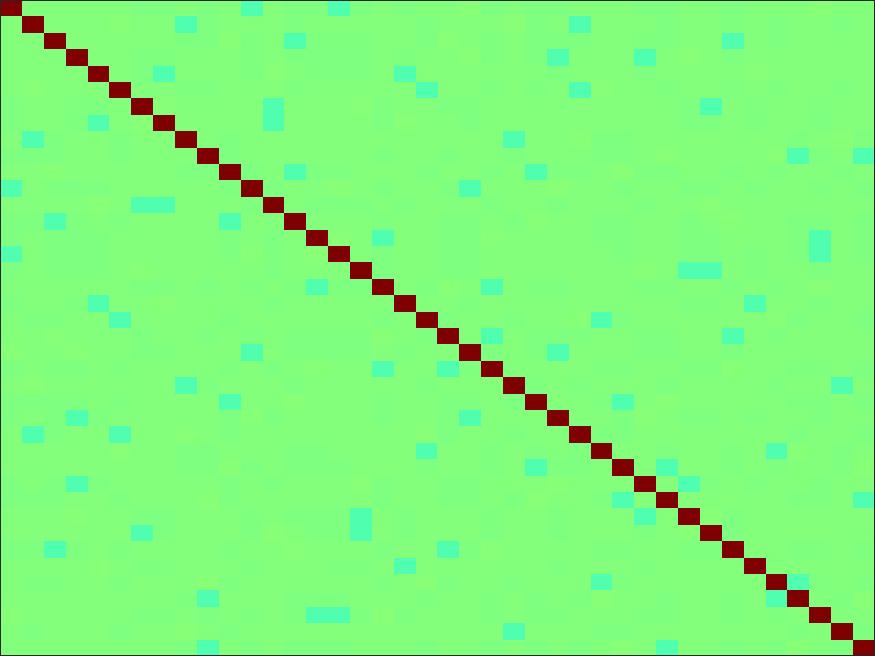}\includegraphics[width=3.5cm,height=3.5cm]{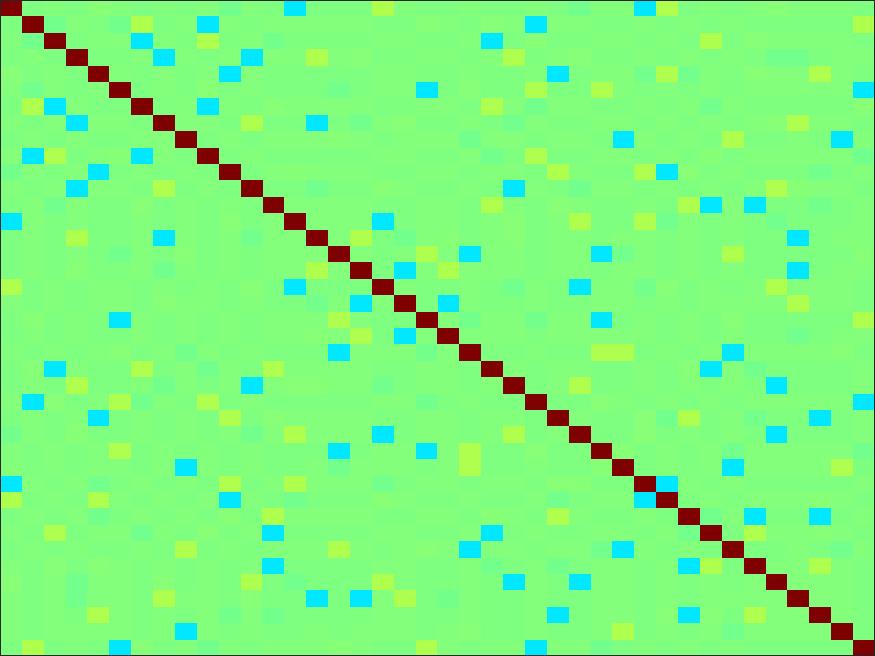}\includegraphics[width=3.5cm,height=3.5cm]{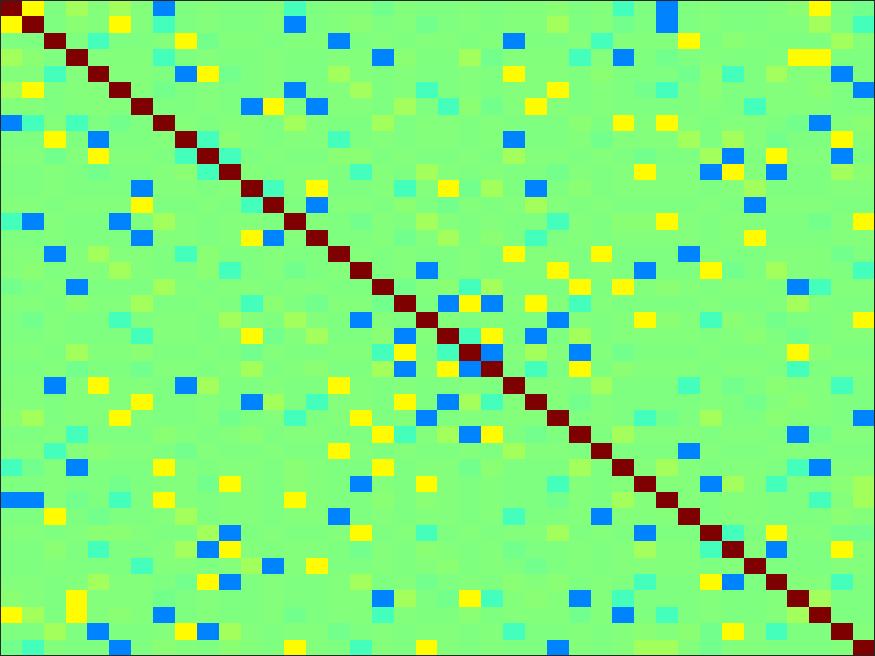}\includegraphics[width=3.5cm,height=3.5cm]{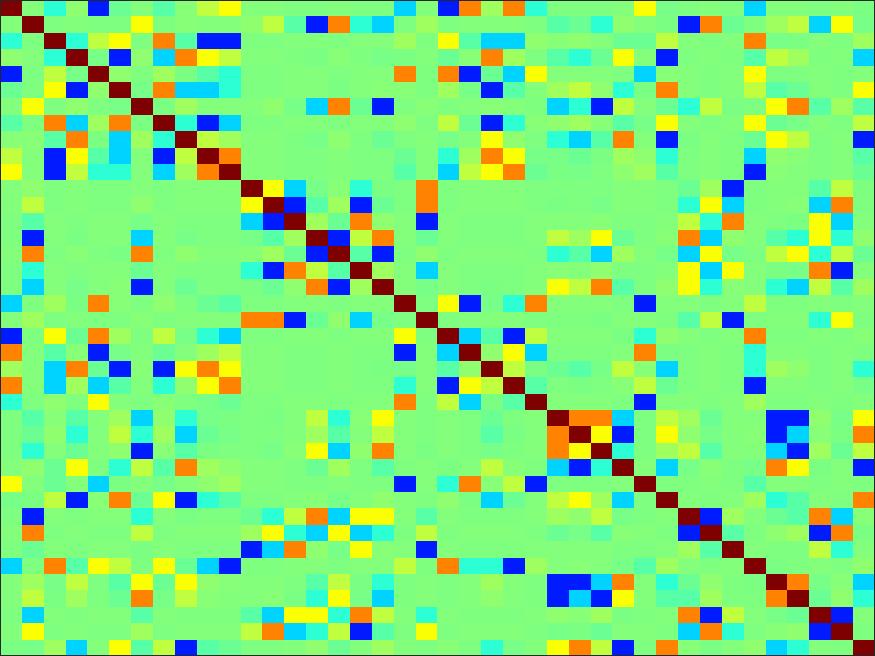}\includegraphics[width=3.5cm,height=3.5cm]{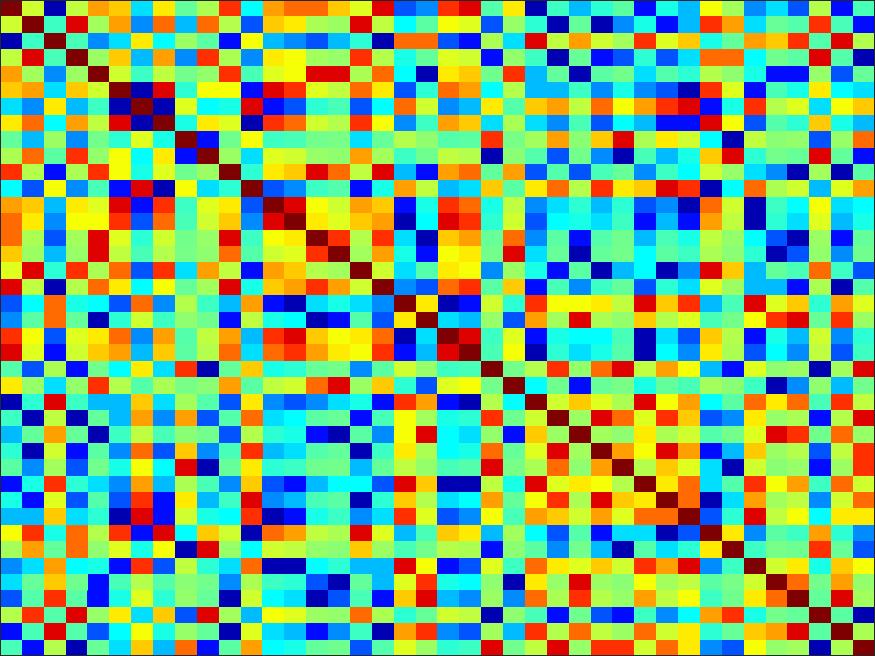}\includegraphics[width=3.5cm,height=3.5cm]{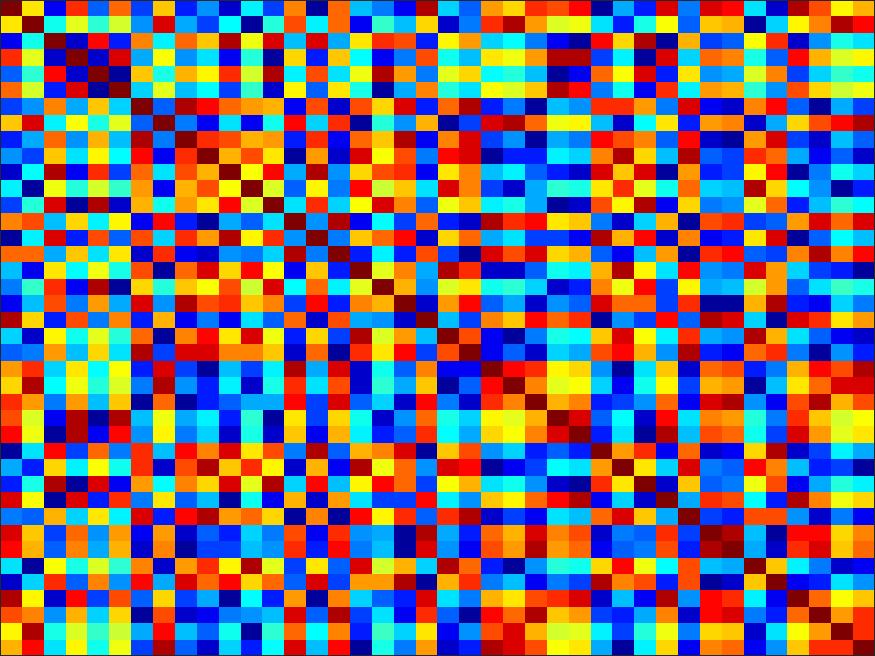}\medskip{}

\begin{centering}
$\ $$\ $$\ $$\ $$\ $\includegraphics[scale=0.62]{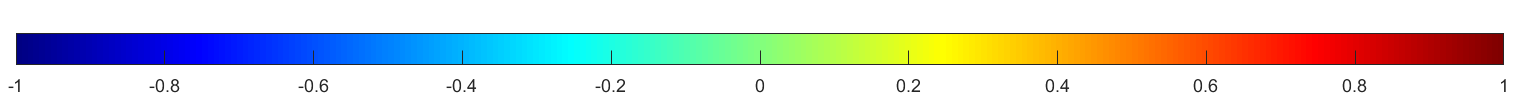}
\par\end{centering}
\caption{Idiosyncratic cross-correlations for $N=200$: (i) Toeplitz matrix with positive $\tau$ (first row); (ii) Permuted columns of Toeplitz matrix with negative $\tau$ (second row).}
\label{fig:Cross}
\end{sidewaysfigure}

Systems of $N=30, 100$ and 200 variables with $T=50, 100$ and 500 observations are generated by model (\ref{eq:DFM}) with $r=1$ and $r=2$ factors, and with the idiosyncratic components, $\varepsilon_t$, generated by a multivariate Gaussian white noise with covariance matrix $\Sigma_{\varepsilon}$. For each design, the number of replications is $R=1000$. We consider two alternative structures for $\Sigma_{\varepsilon}$:
\begin{enumerate}
\item[\textit{Structure} 1] The elements in the main diagonal of $\Sigma_{\varepsilon}$ are generated by $\sigma_i^2\sim U(0.5, 10)$ while the off-diagonal elements are generated according to the same Toeplitz structure described in (\ref{eq:threshold}); see Figure \ref{fig:Cross} for an example of the covariances used to simulate $\varepsilon_t$ when $N=200$ for different values of $\tau$.\footnote{Regardless of $\tau$, the condition of weak idiosyncratic cross-correlation is satisfied. Note that the largest eigenvalue of $\Sigma_{\varepsilon}$ is bounded by the largest absolute row sum, i.e. $l_{max}(\Sigma_{\varepsilon})\leq \max_{j\in (1,...,N)} \sum_{i=1}^N \mid \sigma_{ij}\mid = \max_{j\in (1,...,N)} \sum_{i=1}^N \mid \tau^{\mid i-j \mid} \mid \sigma_{i} \sigma_{j}$. Without loss of generality, we assume that the maximum row sum is associated with the first row. Then, $\sigma_1 \sum_{i=1}^N \mid \tau^{\mid i-1 \mid} \mid \sigma_{i} \leq \sigma_1 \sum_{i=1}^N \mid \tau^{\mid i-j \mid} \mid$. Therefore, $lim_{N \rightarrow \infty l_{max}(\Sigma_{\varepsilon})}\leq \frac{\sigma_1}{1-\mid \tau \mid}.$.} Note that when $\tau=0.9$ or 0.95, although the weak stationarity of the idiosyncratic components is still satisfied, the correlations between the idiosyncratic components are so strong that a further common factor can be appropriate. The cases more realistic from an empirical point of view are those for which $\tau\leq 0.7$. However, we still consider $\tau=0.9$ and 0.95 in our simulations for completeness. 
\item[\textit{Structure} 2] The covariances are generated by a Toeplitz matrix with parameter $\tau=-0.3, -0.5, -0.7, -0.9$ and $-0.95$. The columns are then permuted; see Figure \ref{fig:Cross} for the covariances used to simulate $\varepsilon$ according to this second structure when $N=200$. As before, the correlations for the cases, $\tau=-0.9$ and $-0.95$ suggest the presence of a further common factor in the model. The cases more realistic from an empirical point of view are those for which $|\tau|\leq 0.7$.
\end{enumerate}

With respect to loadings and factors, when $r=1$, the former are generated by $\lambda_{i}\sim U(0,1)$, while the latter is generated by an AR(1) model with the disturbance given by $u_t\sim N(0, 1-\phi^2)$ with $\phi=0.7$.\footnote{We also generate factors with $\phi=0.2$ and 0.95. The results, available upon request, are almost identical to those reported in this section.} Finally, when the number of common factors is $r=2$, they are generated by the following VAR(1) model
\begin{equation}
\label{eq:Factors}
F_{t}=\Phi F_{t-1}+u_{t},
\end{equation}
where $\Phi= diag(0.7, 0.4)$ and $u_{t}$ is an $2\times1$ white noise vector process with variance-covariance matrix given by $\Sigma_{u}= diag (\frac{1}{1-0.7^2}, \frac{1}{1-0.4^2})$. The loadings are generated by two independent $U(0,1)$ distributions. Note that the simulated factors are standardized to satisfy the usual identifiability restrictions, namely, $\frac{F^{\prime}F}{T}=I_{r}$, while the loadings of the second factor are transformed to satisfy $\Lambda^{\prime}\Lambda$ being diagonal.\footnote{In fact, if $\lambda^*_{2i}$ are the $U(0,1)$ loadings generated for the second factor, we obtain $\lambda_{2i}= \lambda_{1i}- \left(\lambda_1^{\prime} \lambda_1 \right) ^{-1} \lambda_1^{\prime} \lambda^*_{2} \lambda_{2i}^*$, where $\lambda_1$ is the $N \times 1$ vector of loadings of the first factor and $\lambda_2^*$ is the $N \times 1$ vector of loadings generated for the second factor.}

Before extracting the factors, the simulated variables are, as usual, centred and standardized.\footnote{It is important to note that by standardizing, the heteroscedasticity is reduced and the simulated systems can be considered as approximately homoscedastic; see the discussion by Poncela and Ruiz (2023).} The factors are extracted by PC and their confidence regions/intervals are constructed by using the asymptotic distribution with the MSE computed using the novel estimator proposed in this paper, AT-CSR, and its corresponding  subsampling correction, AT-CSR*. The empirical coverage of 95\% confidence regions/intervals for the factors is calculated. For comparison purposes, we also compute the asymptotic confidence regions/intervals using the MSE calculated with the HR, CS-HAC1, CS-HAC2, and AV-SHAC estimators, their corresponding subsampling corrections, as well as bootstrap confidence regions/intervals.


\subsection{Performance of confidence intervals: one single factor}

Consider first the DFM with a single factor, i.e. $r=1$. Table \ref{tab:MC5} reports Monte Carlo averages of the empirical coverages of the confidence intervals obtained using the asymptotic distribution with the MSE estimated using the novel AT-CSR estimator proposed in this paper. The results reported in Table \ref{tab:MC5} correspond to the idiosyncratic covariance matrix generated by a Toeplitz matrix with $\tau=0.3, 0.5$ and $0.7$, and by permuting the columns of a Toeplitz matrix with $\tau=-0.3, -0.5$ and $-0.7$.\footnote{Results for other structures of the idiosyncratic cross-correlations are available upon request.} For the sake of comparison, we also consider the case of cross-sectionally uncorrelated idiosyncratic components, i.e. $\tau=0$. In order to analyse the role of the threshold level, $\delta$, in the finite sample properties of the coverages of the intervals for the factor, we consider $\delta=2, 1.7$ and $1.4$ (recall that $\delta=2$ represents a highly sparse idiosyncratic covariance matrix with the sparsity decreasing with $\delta$). 

\begin{table}[H]
{
\centering
\begin{spacing}{0.75}
\begin{tabular}{c|ccc|ccc|ccc}
\hline 
\textbf{\scriptsize{}$N$} &  & \textbf{\scriptsize{}30} &  &  & \textbf{\scriptsize{}100} &  &  & \textbf{\scriptsize{}200} & \tabularnewline
\hline 
\textbf{\scriptsize{}$T$} & \textbf{\scriptsize{}50} & \textbf{\scriptsize{}100} & \textbf{\scriptsize{}500} & \textbf{\scriptsize{}50} & \textbf{\scriptsize{}100} & \textbf{\scriptsize{}500} & \textbf{\scriptsize{}50} & \textbf{\scriptsize{}100} & \textbf{\scriptsize{}500}\tabularnewline
\hline 
\textbf{\scriptsize{}$\tau=0$} &  &  &  &  &  &  &  &  & \tabularnewline
\textbf{\scriptsize{}$\delta=2$} & {\scriptsize{}0.850} & {\scriptsize{}0.894} & {\scriptsize{}0.921} & {\scriptsize{}0.879} & {\scriptsize{}0.914} & {\scriptsize{}0.939} & {\scriptsize{}0.881} & {\scriptsize{}0.918} & {\scriptsize{}0.942}\tabularnewline
\textbf{\scriptsize{}$\delta=1.7$} & {\scriptsize{}0.846} & {\scriptsize{}0.890} & {\scriptsize{}0.911} & {\scriptsize{}0.878} & {\scriptsize{}0.914} & {\scriptsize{}0.939} & {\scriptsize{}0.881} & {\scriptsize{}0.918} & {\scriptsize{}0.942}\tabularnewline
\textbf{\scriptsize{}$\delta=1.4$} & {\scriptsize{}0.829} & {\scriptsize{}0.874} & {\scriptsize{}0.878} & {\scriptsize{}0.873} & {\scriptsize{}0.910} & {\scriptsize{}0.935} & {\scriptsize{}0.878} & {\scriptsize{}0.916} & {\scriptsize{}0.940}\tabularnewline
\hline
\textbf{\scriptsize{}Toeplitz structure} &  &  &  &  &  &  &  &  & \tabularnewline
\hline
\textbf{\scriptsize{}$\tau=0.3$} &  &  &  &  &  &  &  &  & \tabularnewline
\textbf{\scriptsize{}$\delta=2$} & {\scriptsize{}0.745} & {\scriptsize{}0.794} & {\scriptsize{}0.867} & {\scriptsize{}0.784} & {\scriptsize{}0.827} & {\scriptsize{}0.915} & {\scriptsize{}0.792} & {\scriptsize{}0.832} & {\scriptsize{}0.922} \tabularnewline
\textbf{\scriptsize{}$\delta=1.7$} & {\scriptsize{}0.743} & {\scriptsize{}0.797} & {\scriptsize{}0.846} & {\scriptsize{}0.786} & {\scriptsize{}0.836} & {\scriptsize{}0.915} & {\scriptsize{}0.793} & {\scriptsize{}0.840} & {\scriptsize{}0.923} \tabularnewline 
\textbf{\scriptsize{}$\delta=1.4$} & {\scriptsize{}0.731} & {\scriptsize{}0.788} & {\scriptsize{}0.789} & {\scriptsize{}0.790} & {\scriptsize{}0.852} & {\scriptsize{}0.911} & {\scriptsize{}0.797} & {\scriptsize{}0.858} & {\scriptsize{}0.922} \tabularnewline
\hline 
\textbf{\scriptsize{}$\tau=0.5$} &  &  &  &  &  &  &  &  & \tabularnewline
\textbf{\scriptsize{}$\delta=2$} & {\scriptsize{}0.644} & {\scriptsize{}0.745} & {\scriptsize{}0.772} & {\scriptsize{}0.689} & {\scriptsize{}0.797} & {\scriptsize{}0.896} & {\scriptsize{}0.696} & {\scriptsize{}0.797} & {\scriptsize{}0.907} \tabularnewline
\textbf{\scriptsize{}$\delta=1.7$} & {\scriptsize{}0.665} & {\scriptsize{}0.754} & {\scriptsize{}0.720} & {\scriptsize{}0.716} & {\scriptsize{}0.826} & {\scriptsize{}0.900} & {\scriptsize{}0.718} & {\scriptsize{}0.833} & {\scriptsize{}0.912} \tabularnewline 
\textbf{\scriptsize{}$\delta=1.4$} & {\scriptsize{}0.671} & {\scriptsize{}0.730} & {\scriptsize{}0.619} & {\scriptsize{}0.757} & {\scriptsize{}0.840} & {\scriptsize{}0.895} & {\scriptsize{}0.763} & {\scriptsize{}0.855} & {\scriptsize{}0.914} \tabularnewline 
\hline 
\textbf{\scriptsize{}$\tau=0.7$} &  &  &  &  &  &  &  &  & \tabularnewline
\textbf{\scriptsize{}$\delta=2$} & {\scriptsize{}0.560} & {\scriptsize{}0.654} & {\scriptsize{}0.450} & {\scriptsize{}0.600} & {\scriptsize{}0.763} & {\scriptsize{}0.868} & {\scriptsize{}0.599} & {\scriptsize{}0.778} & {\scriptsize{}0.893} \tabularnewline
\textbf{\scriptsize{}$\delta=1.7$} & {\scriptsize{}0.586} & {\scriptsize{}0.633} & {\scriptsize{}0.370} & {\scriptsize{}0.668} & {\scriptsize{}0.789} & {\scriptsize{}0.867} & {\scriptsize{}0.669} & {\scriptsize{}0.809} & {\scriptsize{}0.893} \tabularnewline 
\textbf{\scriptsize{}$\delta=1.4$} & {\scriptsize{}0.574} & {\scriptsize{}0.571} & {\scriptsize{}0.272} & {\scriptsize{}0.720} & {\scriptsize{}0.804} & {\scriptsize{}0.848} & {\scriptsize{}0.740} & {\scriptsize{}0.834} & {\scriptsize{}0.899} \tabularnewline 
\hline
\textbf{\scriptsize{}Permuted correlations} &  &  &  &  &  &  &  &  & \tabularnewline
\hline
\textbf{\scriptsize{}$\tau=-0.3$} &  &  &  &  &  &  &  &  & \tabularnewline
\textbf{\scriptsize{}$\delta=2$} & {\scriptsize{}0.841} & {\scriptsize{}0.890} & {\scriptsize{}0.915} & {\scriptsize{}0.891} & {\scriptsize{}0.929} & {\scriptsize{}0.931} & {\scriptsize{}0.920} & {\scriptsize{}0.955} & {\scriptsize{}0.921} \tabularnewline
\textbf{\scriptsize{}$\delta=1.7$} & {\scriptsize{}0.838} & {\scriptsize{}0.885} & {\scriptsize{}0.902} & {\scriptsize{}0.889} & {\scriptsize{}0.924} & {\scriptsize{}0.931} & {\scriptsize{}0.919} & {\scriptsize{}0.949} & {\scriptsize{}0.921} \tabularnewline 
\textbf{\scriptsize{}$\delta=1.4$} & {\scriptsize{}0.820} & {\scriptsize{}0.866} & {\scriptsize{}0.867} & {\scriptsize{}0.881} & {\scriptsize{}0.911} & {\scriptsize{}0.927} & {\scriptsize{}0.911} & {\scriptsize{}0.933} & {\scriptsize{}0.922} \tabularnewline 
\hline 
\textbf{\scriptsize{}$\tau=-0.5$} &  &  &  &  &  &  &  &  & \tabularnewline
\textbf{\scriptsize{}$\delta=2$} & {\scriptsize{}0.820} & {\scriptsize{}0.871} & {\scriptsize{}0.905} & {\scriptsize{}0.877} & {\scriptsize{}0.901} & {\scriptsize{}0.939} & {\scriptsize{}0.914} & {\scriptsize{}0.922} & {\scriptsize{}0.949} \tabularnewline
\textbf{\scriptsize{}$\delta=1.7$} & {\scriptsize{}0.816} & {\scriptsize{}0.866} & {\scriptsize{}0.889} & {\scriptsize{}0.870} & {\scriptsize{}0.892} & {\scriptsize{}0.939} & {\scriptsize{}0.903} & {\scriptsize{}0.884} & {\scriptsize{}0.953} \tabularnewline 
\textbf{\scriptsize{}$\delta=1.4$} & {\scriptsize{}0.798} & {\scriptsize{}0.847} & {\scriptsize{}0.847} & {\scriptsize{}0.854} & {\scriptsize{}0.888} & {\scriptsize{}0.931} & {\scriptsize{}0.873} & {\scriptsize{}0.873} & {\scriptsize{}0.945} \tabularnewline 
\hline 
\textbf{\scriptsize{}$\tau=-0.7$} &  &  &  &  &  &  &  &  & \tabularnewline
\textbf{\scriptsize{}$\delta=2$} & {\scriptsize{}0.766} & {\scriptsize{}0.829} & {\scriptsize{}0.837} & {\scriptsize{}0.806} & {\scriptsize{}0.875} & {\scriptsize{}0.927} & {\scriptsize{}0.821} & {\scriptsize{}0.868} & {\scriptsize{}0.935} \tabularnewline
\textbf{\scriptsize{}$\delta=1.7$} & {\scriptsize{}0.766} & {\scriptsize{}0.821} & {\scriptsize{}0.805} & {\scriptsize{}0.807} & {\scriptsize{}0.885} & {\scriptsize{}0.924} & {\scriptsize{}0.809} & {\scriptsize{}0.891} & {\scriptsize{}0.935} \tabularnewline 
\textbf{\scriptsize{}$\delta=1.4$} & {\scriptsize{}0.749} & {\scriptsize{}0.792} & {\scriptsize{}0.742} & {\scriptsize{}0.816} & {\scriptsize{}0.882} & {\scriptsize{}0.914} & {\scriptsize{}0.819} & {\scriptsize{}0.897} & {\scriptsize{}0.928} \tabularnewline
\hline 
\end{tabular}
\end{spacing}
}
\caption{Monte Carlo empirical coverage of confidence intervals with nominal coverage 95\% for PC factors constructed using the asymptotic distribution with the MSE estimated using AT-CSR estimator, when the idiosyncratic covariances are generated with a Toeplitz structure with parameter $\tau>0$ and by permuted columns of a Toeplitz matrix with $\tau<0$.}
\label{tab:MC5}
\end{table}

When looking at the Monte Carlo averages of the empirical coverages reported in Table \ref{tab:MC5} obtained when there is not cross-sectional idiosyncratic correlation, i.e. $\tau=0$, we can observe that they are below the nominal except when both $N$ and $T$ are large. This undercoverage could be partially attributed to the failure of the asymptotic MSE of incorporating the uncertainty due to estimation of the loadings. Maldonado and Ruiz (2021) report similar coverages when the asymptotic covariance is estimated by correctly assuming that the idiosyncratic components are cross-sectionally uncorrelated. Therefore, it seems that there is not a price to pay for estimating the MSE of the factors by using the AT-CSR estimator instead of using the estimator that assumes that there is not cross-sectional correlations, when this assumption is true. In terms of the role of the threshold level, the coverages are very similar regardless of $\delta$ and slightly better when $\delta=2$, which could be expected as the true idiosyncratic covariance matrix is diagonal.

Consider now the results when the idiosyncratic components are cross-sectionally correlated. First of all, we can observe that the Monte Carlo averages of the coverages are smaller than when $\tau=0$. However, for large $N$ and $T$, the coverages are very close to the nominal 95\%. Furthermore, with respect to the degree of sparsity to be assumed when implementing the AT-CSR estimator in the presence of idiosyncratic cross-sectional correlation, i.e. $\delta$, Table \ref{tab:MC5} shows that choosing $\delta=2$ works nearly always best when the cross-correlations are generated by the realistic permuted correlations. However, when they are generated by the Toeplitz matrix with large $\tau>0$, one can incur in very mild losses of coverage by choosing $\delta=2$ instead of smaller values, which imply less sparse correlations. Consequently, in this paper, we propose to assume a highly sparse matrix and fix $\delta=2$; see also Qiu and Liyanage (2019) for the same recommendation.

\begin{sidewaysfigure}
\begin{centering}
\includegraphics[scale=0.23]{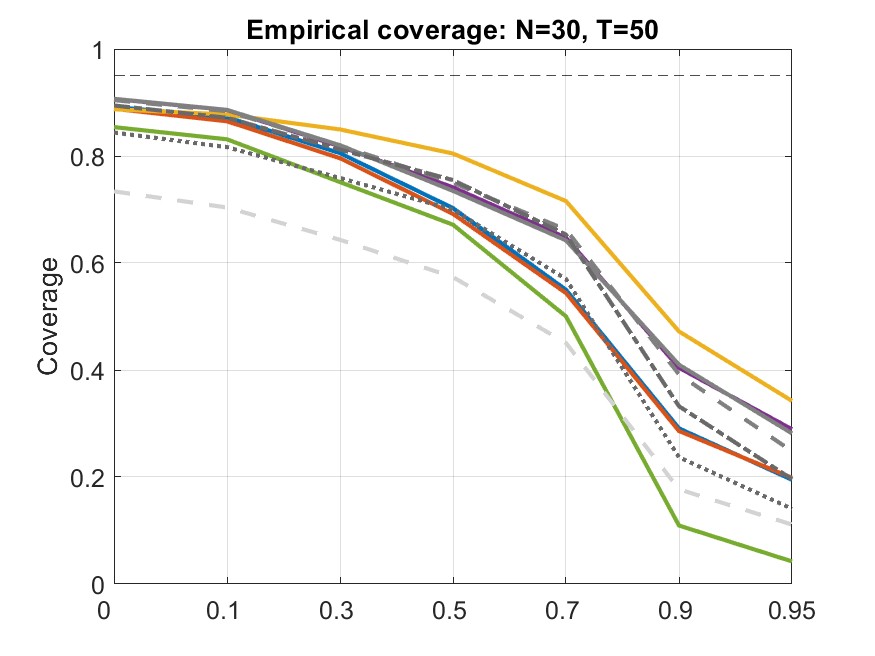}\includegraphics[scale=0.23]{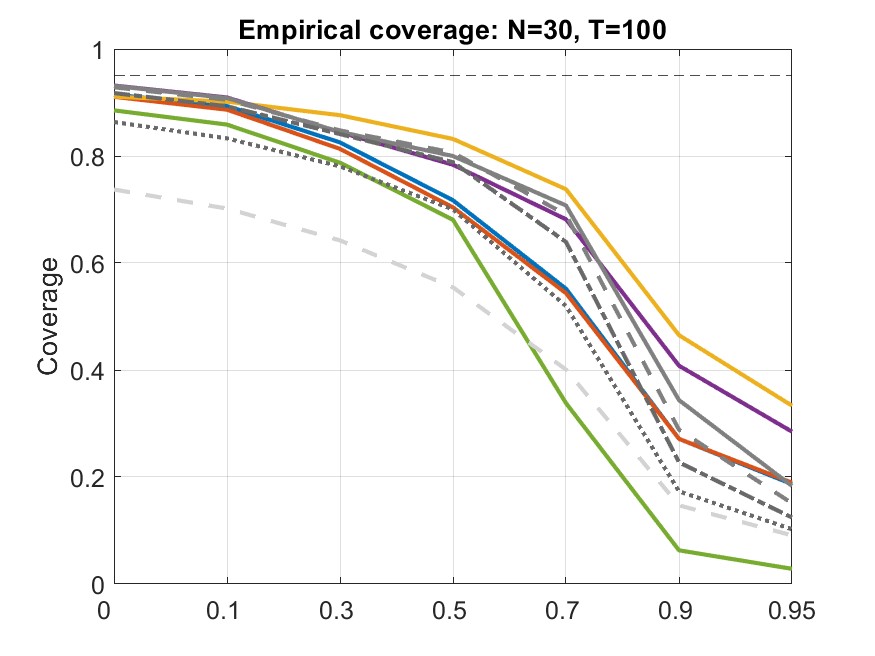}\includegraphics[scale=0.23]{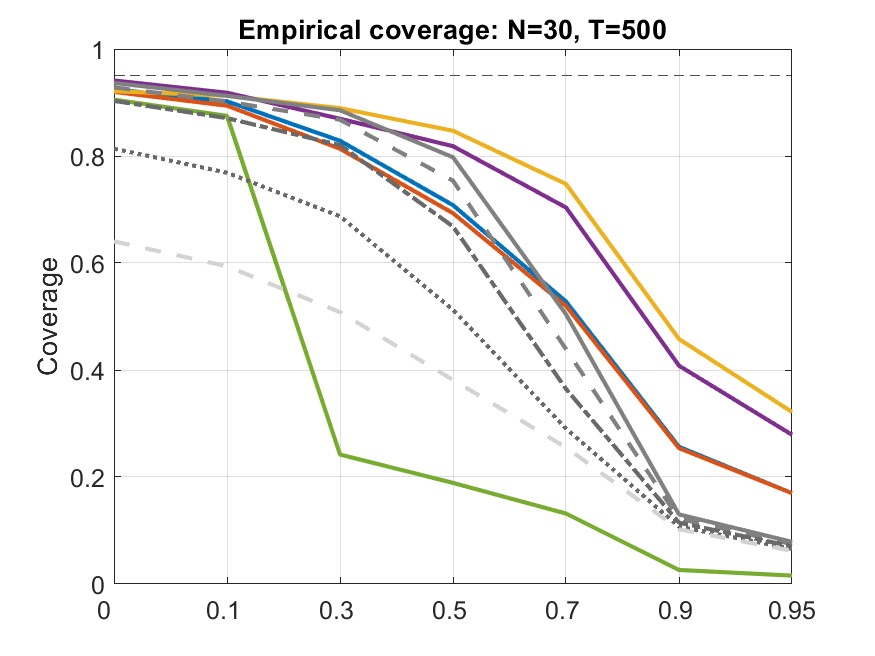}
\par\end{centering}
\begin{centering}
\includegraphics[scale=0.23]{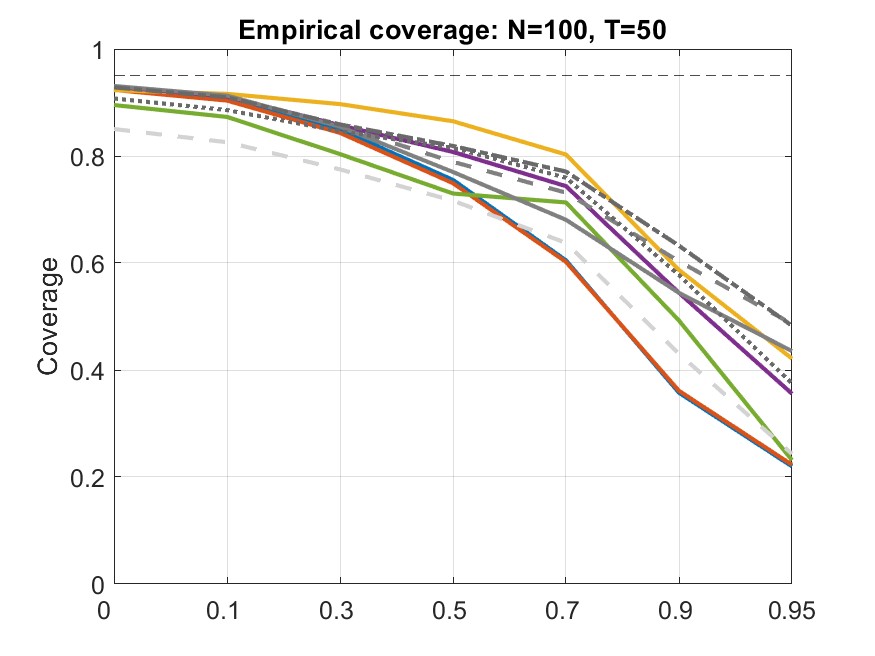}\includegraphics[scale=0.23]{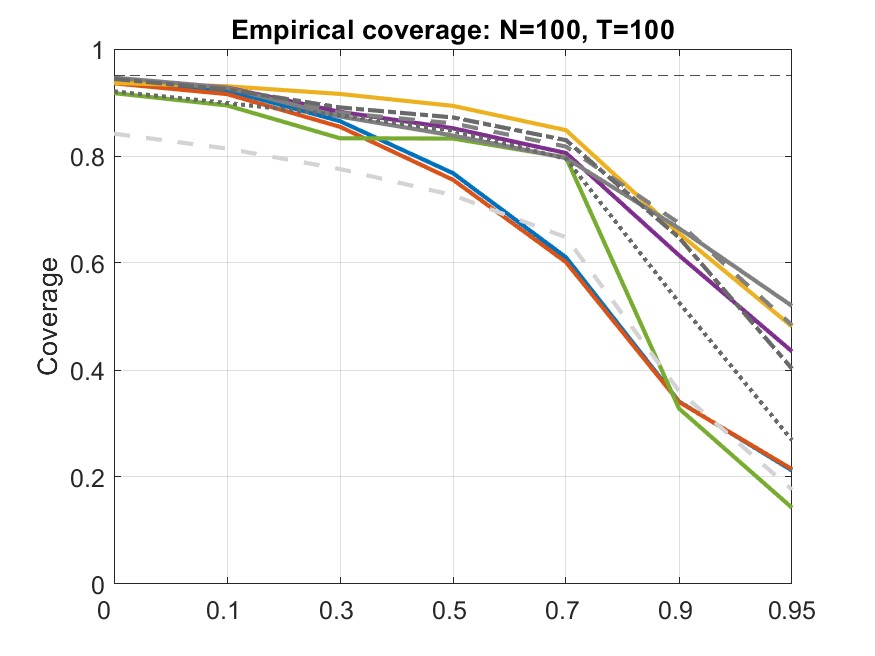}\includegraphics[scale=0.23]{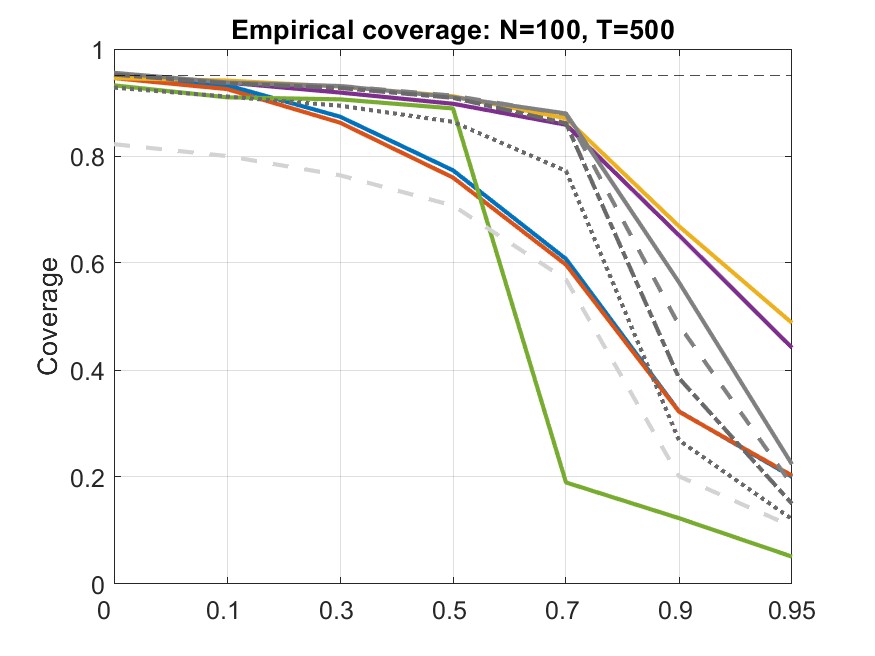}
\par\end{centering}
\begin{centering}
\includegraphics[scale=0.23]{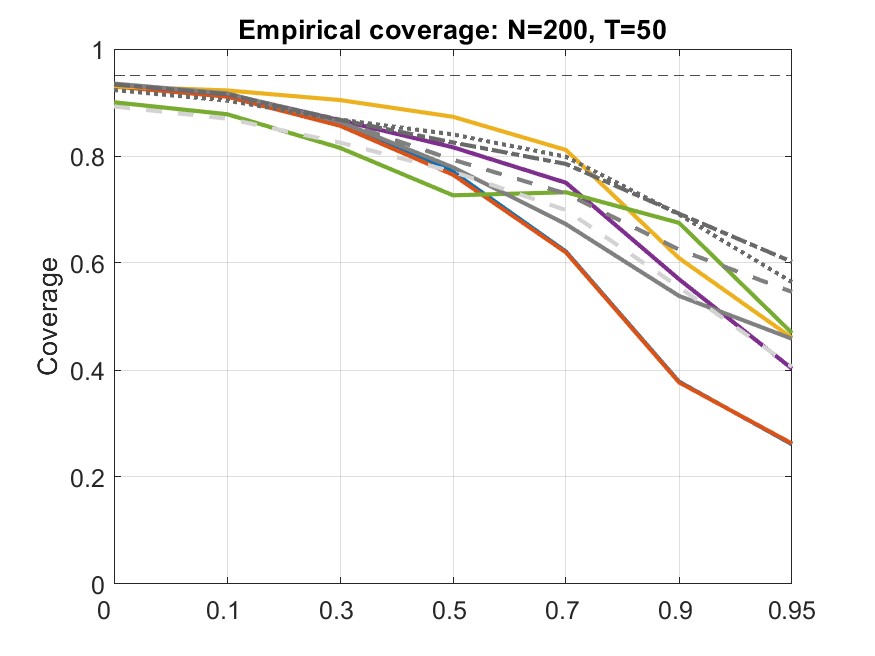}\includegraphics[scale=0.23]{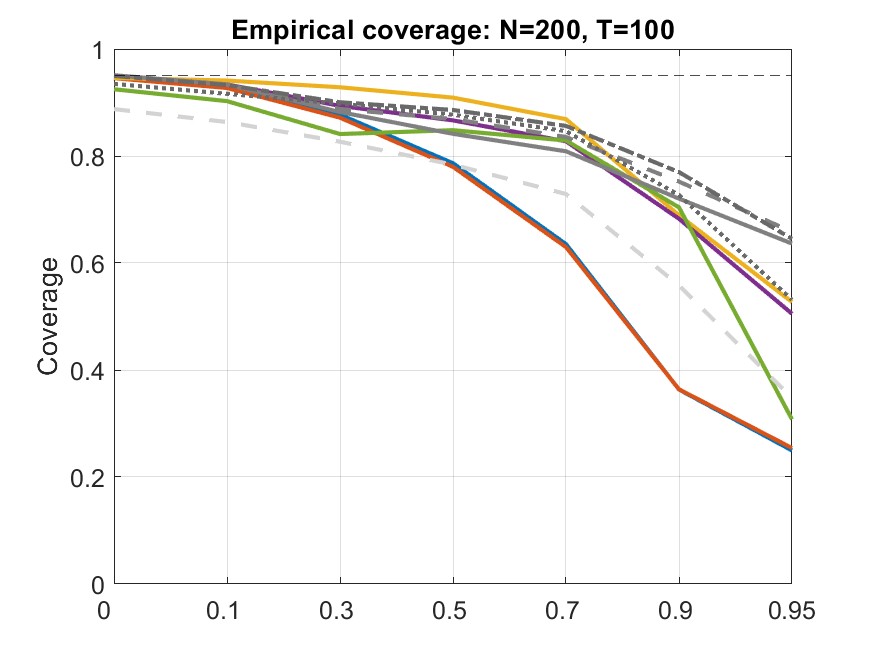}\includegraphics[scale=0.23]{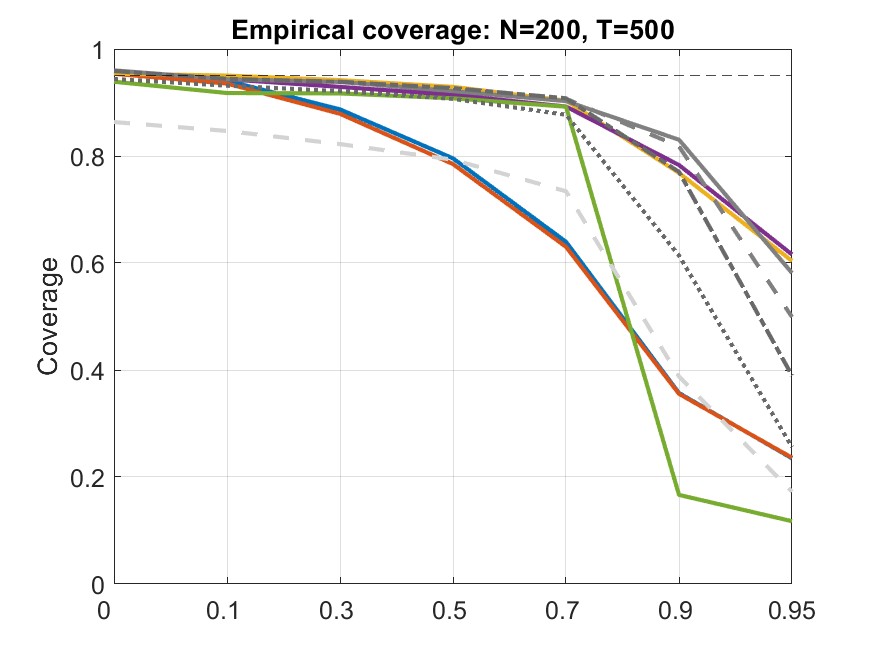}
\par\end{centering}
\caption{Monte Carlo 95\% empirical coverages for confidence intervals for the factor constructed using HR$^{\ast}$ (blue), CS-HAC-1$^{\ast}$ (red), CS-HAC-2$^{\ast}$ (yellow), CSD (green) AV-SHAC$^{\ast}$ (purple), AT-CSR$^{\ast}$ with $\delta=$ 2 (gray), 1.7 (dashed gray), 1.4 (dash-dotted gray), 1 (dotted gray) and 0.5 (dashed lightgray). The DFM has a single factor and the covariances of the idiosyncratic components are generated by a Toeplitz matrix.}
\label{fig:coverages}
\end{sidewaysfigure}

As commented above, the asymptotic MSE of the factors does not incorporate the uncertainty due to the estimation of the loadings. Consequently, we propose using the subsampling correction to incorporate this uncertainty. Table \ref{tab:MC4} reports the empirical coverages when the asymptotic MSE is estimated using the AT-CSR* estimator with $\delta=2$. The idiosyncratic cross-sectional correlations are generated by the Toeplitz matrix. First, observe that, for all $N$ and $T$, the coverages reported in Table \ref{tab:MC4}, when the asymptotic MSE is corrected by subsampling, are very close to the nominal 95\% when $\tau=0$, illustrating the validity of subsampling to incorporate the uncertainty associated with the estimation of the loadings. Second, Table \ref{tab:MC4} also illustrates the asymptotic validity of the proposed AT-CSR* estimator of the MSE of the factors. We can observe that, when $N=200$ and $T=500$, the coverages are close to $95\%$ regardless of the degree of idiosyncratic cross-correlations, $\tau$; see also Figure \ref{fig:coverages} that summarizes the results by plotting the coverages for other values of $\tau$. Third, the confidence intervals based on the proposed estimator of the MSE of the factors show to have good coverages even when $N$ and $T$ are moderate and the level of cross-correlation is large.  

\begin{table}[H]
{
\centering
\begin{spacing}{0.75}
\begin{tabular}{c|ccc|ccc|ccc}
\hline 
\textbf{\scriptsize{}$N$} &  & \textbf{\scriptsize{}30} &  &  & \textbf{\scriptsize{}100} &  &  & \textbf{\scriptsize{}200} & \tabularnewline
\hline 
\textbf{\scriptsize{}$T$} & \textbf{\scriptsize{}50} & \textbf{\scriptsize{}100} & \textbf{\scriptsize{}500} & \textbf{\scriptsize{}50} & \textbf{\scriptsize{}100} & \textbf{\scriptsize{}500} & \textbf{\scriptsize{}50} & \textbf{\scriptsize{}100} & \textbf{\scriptsize{}500}\tabularnewline
\hline 
\textbf{\scriptsize{}$\tau=0$} &  &  &  &  &  &  &  &  & \tabularnewline
\textbf{\scriptsize{}Asymptotic} &  &  &  &  &  &  &  &  & \tabularnewline
\textbf{\scriptsize{}AT-SCR*} & {\scriptsize{}0.907} & {\scriptsize{}0.931} & {\scriptsize{}0.937} & {\scriptsize{}0.931} & {\scriptsize{}0.946} & {\scriptsize{}0.955} & {\scriptsize{}0.935} & {\scriptsize{}0.951} & {\scriptsize{}0.960}\tabularnewline
\textbf{\scriptsize{}HR*} & {\scriptsize{}0.892} & {\scriptsize{}0.915} & {\scriptsize{}0.925} & {\scriptsize{}0.927} & {\scriptsize{}0.942} & {\scriptsize{}0.951} & {\scriptsize{}0.933} & {\scriptsize{}0.949} & {\scriptsize{}0.958}\tabularnewline
\textbf{\scriptsize{}SC-HAC1*} & {\scriptsize{}0.888} & {\scriptsize{}0.911} & {\scriptsize{}0.920} & {\scriptsize{}0.924} & {\scriptsize{}0.936} & {\scriptsize{}0.946} & {\scriptsize{}0.930} & {\scriptsize{}0.946} & {\scriptsize{}0.954}\tabularnewline
\textbf{\scriptsize{}SC-HAC2*} & {\scriptsize{}0.887} & {\scriptsize{}0.911} & {\scriptsize{}0.921} & {\scriptsize{}0.923} & {\scriptsize{}0.937} & {\scriptsize{}0.946} & {\scriptsize{}0.930} & {\scriptsize{}0.946} & {\scriptsize{}0.954}\tabularnewline
\textbf{\scriptsize{}AV-SHAC*} & {\scriptsize{}0.907} & {\scriptsize{}0.932} & {\scriptsize{}0.941} & {\scriptsize{}0.929} & {\scriptsize{}0.946} & {\scriptsize{}0.955} & {\scriptsize{}0.935} & {\scriptsize{}0.951} & {\scriptsize{}0.960}\tabularnewline
\textbf{\scriptsize{}Bootstrap} &  &  &  &  &  &  &  &  & \tabularnewline
\textbf{\scriptsize{}CSD} & {\scriptsize{}0.854} & {\scriptsize{}0.885} & {\scriptsize{}0.905} & {\scriptsize{}0.895} & {\scriptsize{}0.918} & {\scriptsize{}0.932} & {\scriptsize{}0.900} & {\scriptsize{}0.925} & {\scriptsize{}0.939}\tabularnewline 
\hline
\textbf{\scriptsize{}$\tau=0.3$} &  &  &  &  &  &  &  &  & \tabularnewline
\textbf{\scriptsize{}Asymptotic} &  &  &  &  &  &  &  &  & \tabularnewline
\textbf{\scriptsize{}AT-SCR*} & {\scriptsize{}0.820} & {\scriptsize{}0.845} & {\scriptsize{}0.885} & {\scriptsize{}0.855} & {\scriptsize{}0.874} & {\scriptsize{}0.931} & {\scriptsize{}0.864} & {\scriptsize{}0.882} & {\scriptsize{}0.938} \tabularnewline
\textbf{\scriptsize{}HR*} & {\scriptsize{}0.805} & {\scriptsize{}0.825} & {\scriptsize{}0.828} & {\scriptsize{}0.850} & {\scriptsize{}0.865} & {\scriptsize{}0.874} & {\scriptsize{}0.861} & {\scriptsize{}0.877} & {\scriptsize{}0.887}\tabularnewline
\textbf{\scriptsize{}SC-HAC1*} & {\scriptsize{}0.796} & {\scriptsize{}0.813} & {\scriptsize{}0.814} & {\scriptsize{}0.843} & {\scriptsize{}0.855} & {\scriptsize{}0.862} & {\scriptsize{}0.857} & {\scriptsize{}0.872} & {\scriptsize{}0.879}\tabularnewline
\textbf{\scriptsize{}SC-HAC2*} & {\scriptsize{}0.850} & {\scriptsize{}0.876} & {\scriptsize{}0.889} & {\scriptsize{}0.897} & {\scriptsize{}0.916} & {\scriptsize{}0.930} & {\scriptsize{}0.905} & {\scriptsize{}0.928} & {\scriptsize{}0.942}\tabularnewline
\textbf{\scriptsize{}AV-SHAC*} & {\scriptsize{}0.818} & {\scriptsize{}0.844} & {\scriptsize{}0.870} & {\scriptsize{}0.857} & {\scriptsize{}0.882} & {\scriptsize{}0.919} & {\scriptsize{}0.867} & {\scriptsize{}0.894} & {\scriptsize{}0.929}\tabularnewline
\textbf{\scriptsize{}Bootstrap} &  &  &  &  &  &  &  &  & \tabularnewline
\textbf{\scriptsize{}CSD} & {\scriptsize{}0.751} & {\scriptsize{}0.787} & {\scriptsize{}0.242} & {\scriptsize{}0.804} & {\scriptsize{}0.833} & {\scriptsize{}0.906} & {\scriptsize{}0.815} & {\scriptsize{}0.841} & {\scriptsize{}0.917} \tabularnewline 
\hline 
\textbf{\scriptsize{}$\tau=0.5$} &  &  &  &  &  &  &  &  & \tabularnewline
\textbf{\scriptsize{}Asymptotic} &  &  &  &  &  &  &  &  & \tabularnewline
\textbf{\scriptsize{}AT-SCR*} & {\scriptsize{}0.735} & {\scriptsize{}0.800} & {\scriptsize{}0.798} & {\scriptsize{}0.770} & {\scriptsize{}0.838} & {\scriptsize{}0.910} & {\scriptsize{}0.779} & {\scriptsize{}0.841} & {\scriptsize{}0.921} \tabularnewline
\textbf{\scriptsize{}HR*} & {\scriptsize{}0.703} & {\scriptsize{}0.717} & {\scriptsize{}0.708} & {\scriptsize{}0.755} & {\scriptsize{}0.768} & {\scriptsize{}0.773} & {\scriptsize{}0.770} & {\scriptsize{}0.786} & {\scriptsize{}0.796}\tabularnewline
\textbf{\scriptsize{}SC-HAC1*} & {\scriptsize{}0.692} & {\scriptsize{}0.704} & {\scriptsize{}0.693} & {\scriptsize{}0.749} & {\scriptsize{}0.756} & {\scriptsize{}0.760} & {\scriptsize{}0.765} & {\scriptsize{}0.780} & {\scriptsize{}0.785}\tabularnewline
\textbf{\scriptsize{}SC-HAC2*} & {\scriptsize{}0.805} & {\scriptsize{}0.832} & {\scriptsize{}0.847} & {\scriptsize{}0.865} & {\scriptsize{}0.894} & {\scriptsize{}0.912} & {\scriptsize{}0.873} & {\scriptsize{}0.909} & {\scriptsize{}0.929}\tabularnewline
\textbf{\scriptsize{}AV-SHAC*} & {\scriptsize{}0.741} & {\scriptsize{}0.784} & {\scriptsize{}0.818} & {\scriptsize{}0.808} & {\scriptsize{}0.852} & {\scriptsize{}0.898} & {\scriptsize{}0.816} & {\scriptsize{}0.867} & {\scriptsize{}0.914}\tabularnewline
\textbf{\scriptsize{}Bootstrap} &  &  &  &  &  &  &  &  & \tabularnewline
\textbf{\scriptsize{}CSDA} & {\scriptsize{}0.672} & {\scriptsize{}0.681} & {\scriptsize{}0.189} & {\scriptsize{}0.730} & {\scriptsize{}0.833} & {\scriptsize{}0.889} & {\scriptsize{}0.727} & {\scriptsize{}0.848} & {\scriptsize{}0.908} \tabularnewline
\hline 
\textbf{\scriptsize{}$\tau=0.7$} &  &  &  &  &  &  &  &  & \tabularnewline
\textbf{\scriptsize{}Asymptotic} &  &  &  &  &  &  &  &  & \tabularnewline
\textbf{\scriptsize{}AT-SCR*} & {\scriptsize{}0.642} & {\scriptsize{}0.708} & {\scriptsize{}0.504} & {\scriptsize{}0.681} & {\scriptsize{}0.796} & {\scriptsize{}0.879} & {\scriptsize{}0.672} & {\scriptsize{}0.809} & {\scriptsize{}0.903} \tabularnewline
\textbf{\scriptsize{}HR*} & {\scriptsize{}0.550} & {\scriptsize{}0.552} & {\scriptsize{}0.528} & {\scriptsize{}0.605} & {\scriptsize{}0.610} & {\scriptsize{}0.608} & {\scriptsize{}0.621} & {\scriptsize{}0.635} & {\scriptsize{}0.640}\tabularnewline
\textbf{\scriptsize{}SC-HAC1*} & {\scriptsize{}0.544} & {\scriptsize{}0.544} & {\scriptsize{}0.520} & {\scriptsize{}0.602} & {\scriptsize{}0.602} & {\scriptsize{}0.597} & {\scriptsize{}0.620} & {\scriptsize{}0.630} & {\scriptsize{}0.630}\tabularnewline
\textbf{\scriptsize{}SC-HAC2*} & {\scriptsize{}0.716} & {\scriptsize{}0.738} & {\scriptsize{}0.748} & {\scriptsize{}0.803} & {\scriptsize{}0.848} & {\scriptsize{}0.870} & {\scriptsize{}0.811} & {\scriptsize{}0.869} & {\scriptsize{}0.903}\tabularnewline
\textbf{\scriptsize{}AV-SHAC*} & {\scriptsize{}0.647} & {\scriptsize{}0.682} & {\scriptsize{}0.704} & {\scriptsize{}0.744} & {\scriptsize{}0.806} & {\scriptsize{}0.858} & {\scriptsize{}0.750} & {\scriptsize{}0.827} & {\scriptsize{}0.892}\tabularnewline
\textbf{\scriptsize{}Bootstrap} &  &  &  &  &  &  &  &  & \tabularnewline
\textbf{\scriptsize{}CSD} & {\scriptsize{}0.500} & {\scriptsize{}0.338} & {\scriptsize{}0.132} & {\scriptsize{}0.713} & {\scriptsize{}0.797} & {\scriptsize{}0.190} & {\scriptsize{}0.732} & {\scriptsize{}0.829} & {\scriptsize{}0.892} \tabularnewline
\hline
\end{tabular}
\end{spacing}
}
\caption{Monte Carlo empirical coverage of confidence intervals with nominal coverage 95\% for PC factors constructed using the asymptotic distribution with the MSE estimated by HR*, SC-HAC1*, SC-HAC2*, AV-SHAC*, AT-CSR* with $\delta=2$, and by bootstrapping, CSD, when the idiosyncratic covariances are generated with a Toeplitz structure with $\tau>0$.}
\label{tab:MC4}
\end{table} 

Finally, we compare the coverages of the AT-CSR* estimator with those of extant asymptotic HR*, CS-HAC1*, CS-HAC2*, and AV-SHAC* estimators, and of the bootstrap CSD.\footnote{Results for the estimators of the asymptotic MSE without subsampling correction are available upon request.} Table \ref{tab:MC4} reports the Monte Carlo averages of the coverages for all estimators when the structure of the cross-correlations is generated by a Toeplitz matrix; see also Figure \ref{fig:coverages}. First, we can observe that, when $\tau=0$, intervals based on all estimators of the asymptotic MSE corrected by subsampling have coverages very close to the nominal. We can only observe a slight undercoverage in the bootstrap intervals, which is expected as they are not designed to incorporate the loading estimation uncertainty.
 
Second, if the idiosyncratic components are cross-sectionally correlated, the best coverages are obtained when the infeasible SC-HAC2 estimator is implemented. This is due to the fact that the estimator is assuming the true structure of the correlations. However, if we look at the coverages of the more realistic SC-HAC1 estimator, we can observe that the coverages are similar, and even worse in some cases, to those obtained when using the HR estimator that assumes that the idiosyncratic components are uncorrelated. Both of these two latter estimators show severe undercoverages in some cases. The performance of the bootstrap CSD intervals can be poor, mainly when $T$ is large as compared with $N$. Finally, it is remarkable that the intervals based on the novel AT-SCR estimator have coverages competitive with those of the AV-SHAC estimator, which are clearly better than those of the HR, SC-HAC1 and CSDA estimators. The AT-SCR is a computationally simple alternative to AV-SHAC with similar performance.

\subsection{Performance of confidence regions: two factors}

Consider now the DFM with $r=2$ and the 95\% confidence ellipses constructed by assuming asymptotic normality with the asymptotic MSEs obtained using AT-SCR with $\delta=2$, HR, and AV-SHAC, and their corresponding subsampling corrections. We consider the case in which the cross-correlations of the idiosyncratic components are obtained by permuting the columns of the Toeplitz matrix with negative parameter $\tau$.

\begin{sidewaysfigure}
\begin{centering}
\includegraphics[scale=0.23]{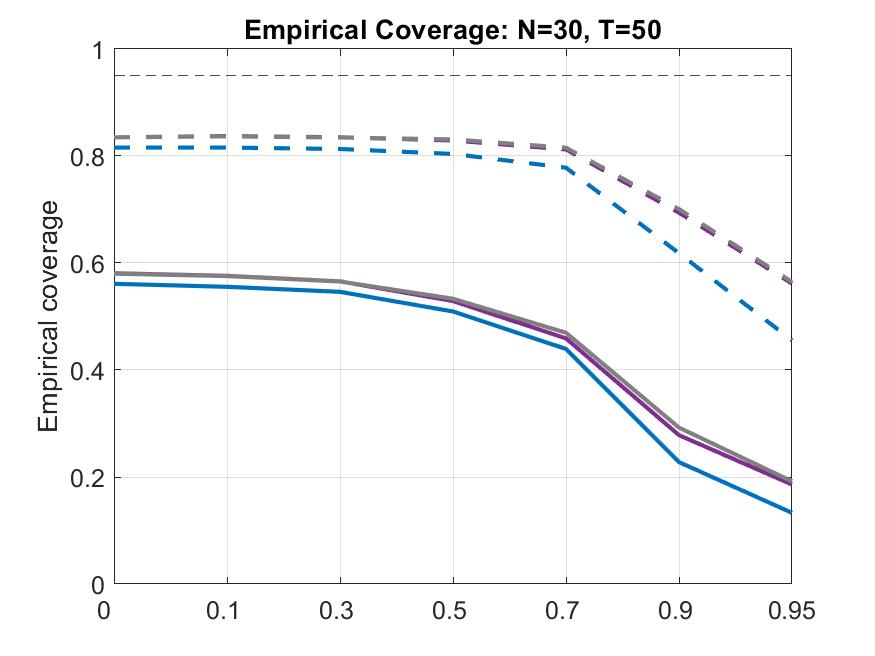}\includegraphics[scale=0.23]{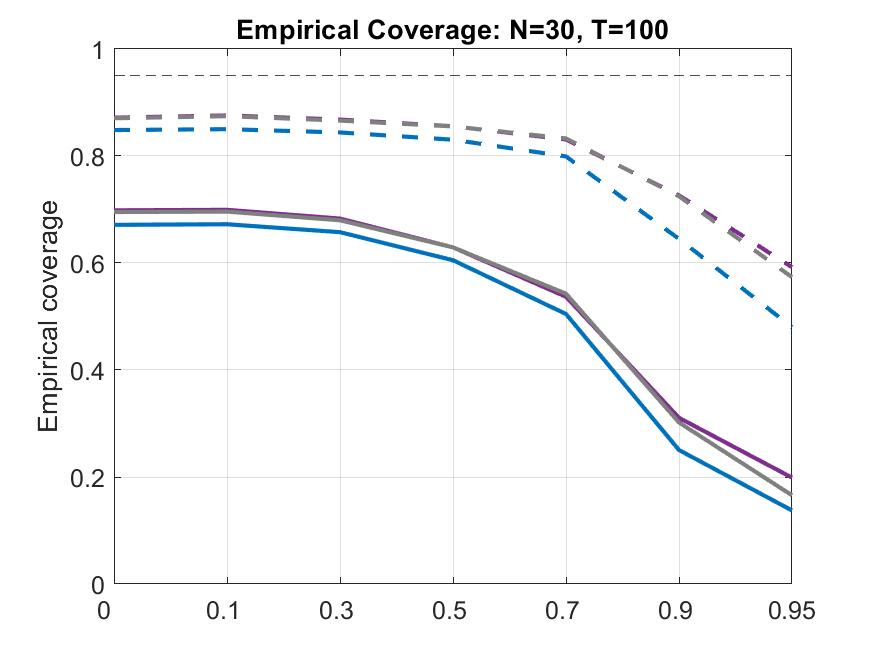}\includegraphics[scale=0.23]{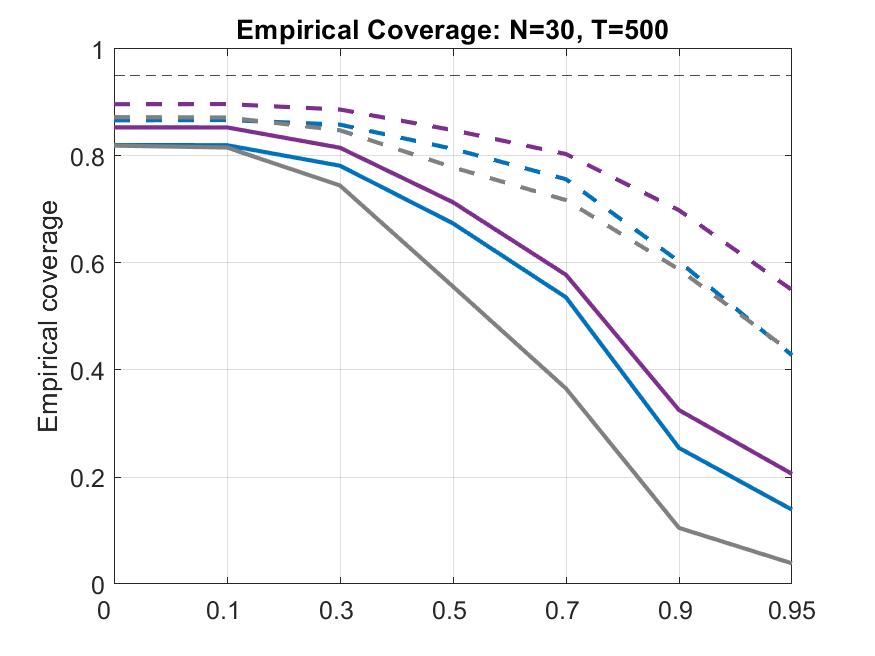}
\par\end{centering}
\begin{centering}
\includegraphics[scale=0.23]{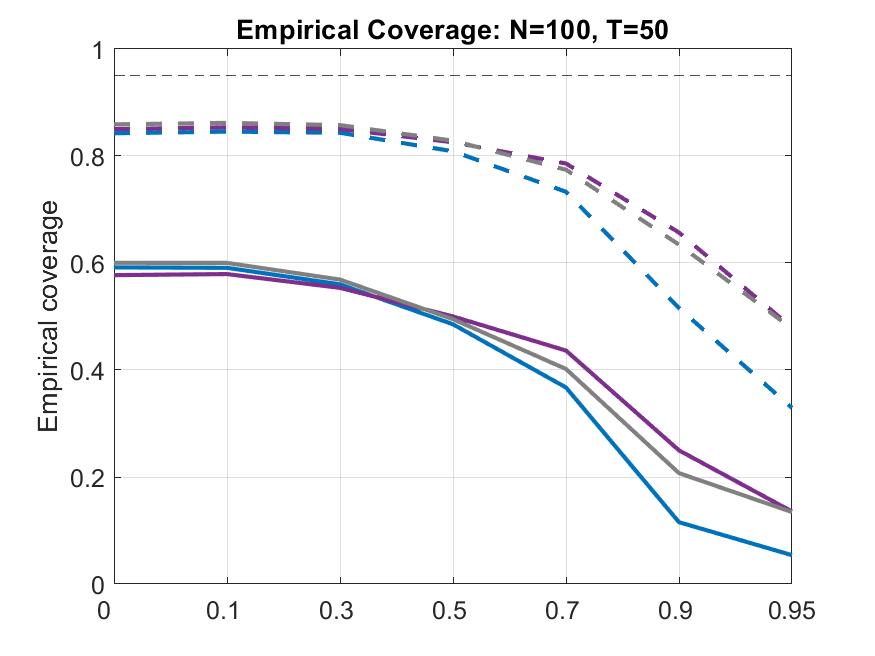}\includegraphics[scale=0.23]{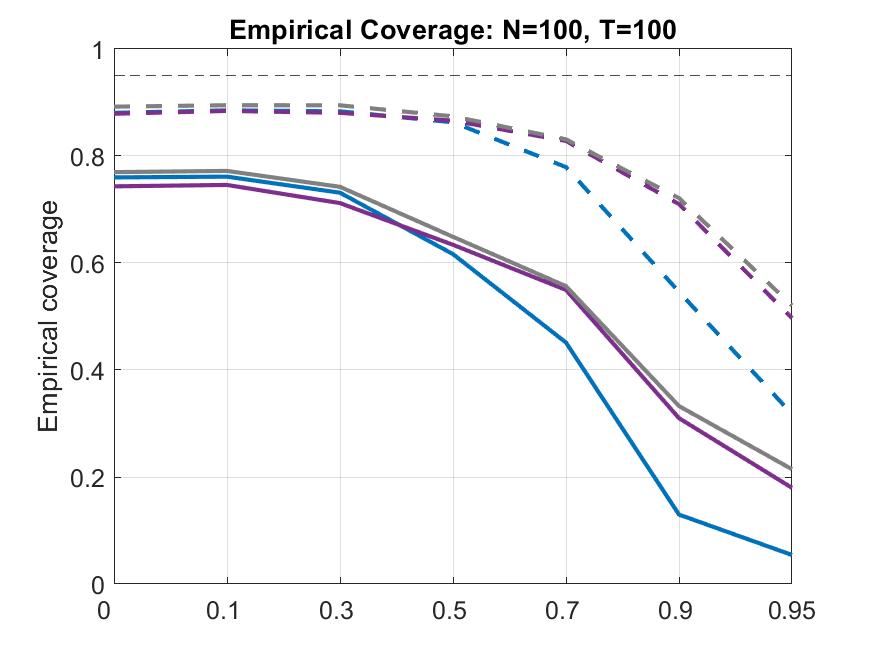}\includegraphics[scale=0.23]{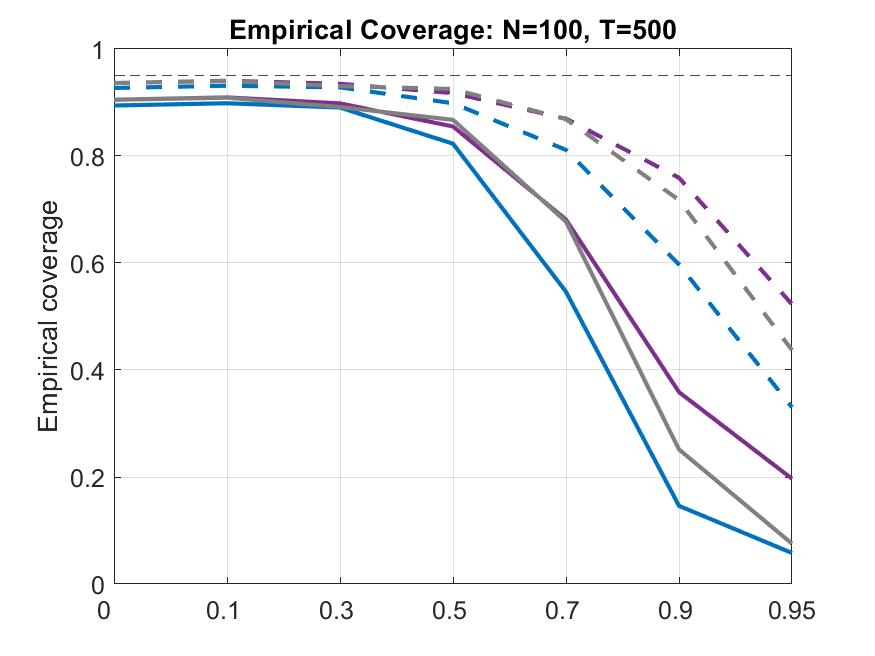}
\par\end{centering}
\begin{centering}
\includegraphics[scale=0.23]{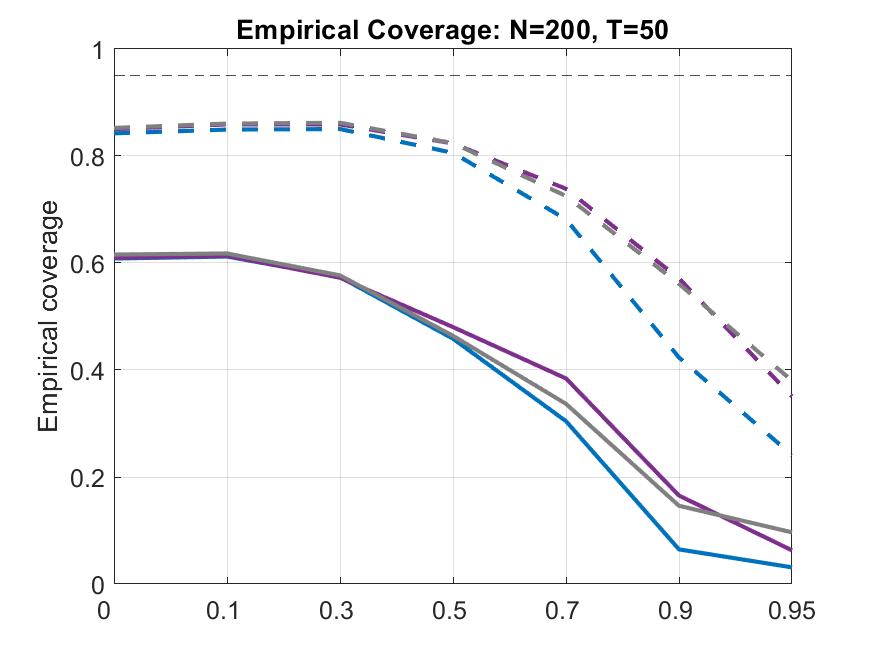}\includegraphics[scale=0.23]{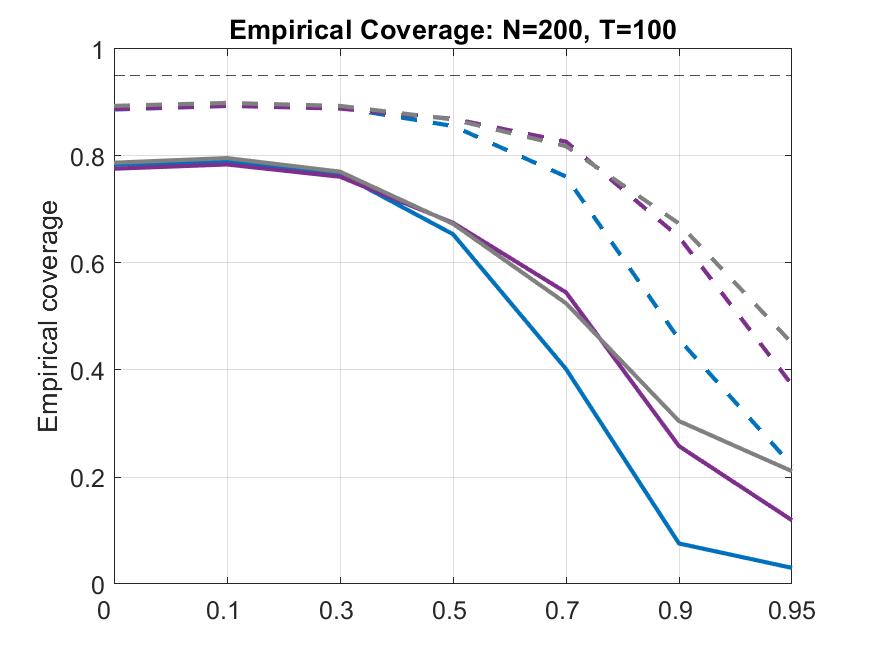}\includegraphics[scale=0.23]{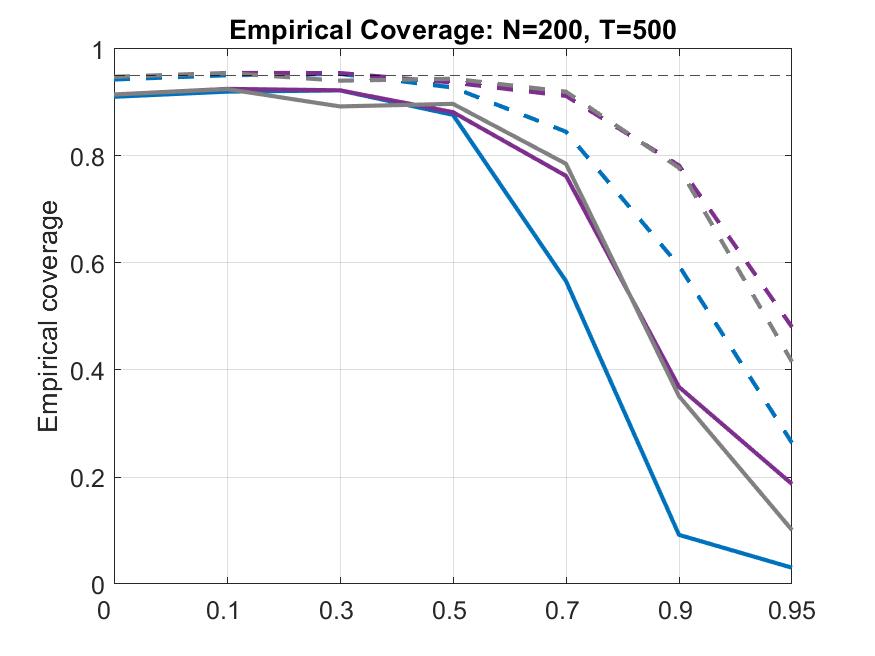}
\par\end{centering}
\caption{Monte Carlo 95\% empirical coverage for Bonferroni confidence regions constructed using HR (blue), AV-HAC (purple), AT-CSR with $\delta=2$ (gray) and their corresponding bootstrap procedures (dashed lines). The DFM has two factors and the covariances of the idiosyncratic components are generated by permuting the columns of a Toeplitz matrix with negative parameter.}
\label{fig:coverages2}
\end{sidewaysfigure}

Figure \ref{fig:coverages2} plots the coverages of the  confidence ellipses for different values of $\tau$. The main conclusions are as above. First, using the threshold level $\delta=2$ guarantees good coverage of the ellipses  for all sample sizes and levels of cross-correlation. Only when $N=30$ and $T=500$, we observe that the performance is slightly worse if the level of cross-correlation is high. In this case, it could be advisable to use a smaller value of $\delta$.  Second, the performance of the computationally simple AT-SCR estimator of the asymptotic MSE of the factors is comparable with that of the estimator proposed by Kim (2022), which requires a computationally demanding procedure to select the bandwidth parameter. It seems that nothing is lost by using the adapting threshold when estimating the uncertainty of the factors in the presence of cross-sectionally correlated idiosyncratic components. Third, using the subsampling to incorporate the uncertainty due to the estimation of the loadings improves the coverages , which are rather close to the nominal. Note that the importance of taking into account the uncertainty of the loadings is even larger than when $r=1$. Finally, Figure \ref{fig:coverages2} also illustrates the asymptotic validity of the proposed AT-SCR* estimator. Regardless of the level of cross-sectional dependence of the idiosyncratic components, the ellipses computed using the AT-SCR* estimator have coverages equal to the nominal if $N=200$ and $T=500$. Only if $\tau>0.9$, the coverages are below the nominal 95\%.

\section{Conclusions}

In this paper, we propose an estimator of the asymptotic MSE of PC factors that is asymptotically valid in the presence of weak cross-correlations of the idiosyncratic components. The novel estimator is based on adaptive thresholding of the sample covariances of the idiosyncratic PC residuals. The estimator is computationally very simple and does not depend on nuisance parameters. We also propose incorporating the uncertainty associated with the estimation of the loadings into the MSE by using a subsampling procedure.

The good performance of the proposed estimator of the MSE of the factors is illustrated by carrying out extensive Monte Carlo simulations. The coverages of the confidence intervals/regions are comparable to those obtained by using the best extant procedures available in the related literature being at the same time less involved computationally.

\end{document}